\documentclass{amsart} 
\usepackage[table]{xcolor}

\usepackage[utf8]{inputenc}
\usepackage[T1]{fontenc}
\usepackage{fourier, erewhon}
\usepackage{geometry}
\usepackage{array, caption, floatrow, tabularx, makecell, booktabs}%
\setcellgapes{3pt}
\usepackage{xcolor,colortbl}
\definecolor{green}{rgb}{0.1,0.1,0.1}

\usepackage{hyperref}

\usepackage{graphicx}  
\graphicspath{{.},{./fig/}} 
\DeclareGraphicsExtensions{.png,.PNG,.pdf,.PDF,.eps,.EPS}

\usepackage{subcaption} 
\usepackage{float}
\usepackage[section]{placeins} 

\usepackage{diagbox}
\usepackage{slashbox}
\usepackage{algorithm}
\usepackage{algpseudocode}

\usepackage{algorithm}
\usepackage{algpseudocode}

\usepackage{mathtools} 
\usepackage{amsthm}

\usepackage{cleveref}  
\usepackage{siunitx} 

\usepackage{tikz}
\usepackage{pgfplots}
\pgfplotsset{compat=1.18}
\pgfplotsset{
    table/search path={./fig/}, 
}

\renewcommand{\vec}{\boldsymbol}

\DeclarePairedDelimiter{\norm}{\Vert}{\Vert}
\DeclarePairedDelimiter{\abs}{\vert}{\vert}
\DeclarePairedDelimiter{\pos}{\lbrack}{\rbrack^+}
\DeclareMathOperator{\rank}{rank}
\DeclareMathOperator{\E}{\mathbb{E}}

\renewcommand{\epsilon}{\varepsilon}

\usepackage[colorinlistoftodos]{todonotes}
\setuptodonotes{size=\scriptsize,backgroundcolor=red!15!white}

\usepackage[mathlines]{lineno}

\usepackage{amsrefs}

\title[]
{
    Detecting transitions between collective motion regimes using functional hypothesis test of the time-varying persistence homology
}

\author[T.~Sathiyakumar et al.]{Thevasha Sathiyakumar}
\email{sathiyt@clarkson.edu}
\address{Department of Mathematics, Clarkson University}
\author{Shantanu Sur}
\email{ssur@clarkson.edu}
\address{Department of Biology, Clarkson University}
\author{Sumona Mondal}
\email{smondal@clarkson.edu}
\author{Marko Budi\v{s}i\'{c}}
\email{mbudisic@clarkson.edu}
\address{Department of Mathematics, Clarkson University}

\begin{document}
\begin{abstract}

In a system of many similar self-propelled entities such as flocks of birds, fish school, cells and molecules, the interactions with neighbors can lead to a "coherent state”, meaning the
formation of visually compelling aggregation patterns due to the local adjustment of speed and direction. In this study, we explore one of the open questions that arise in studying collective patterns. When such entities, considered here as particles, tend to assume a coherent state
beginning from an incoherent (random) state, what is the time interval for the transition? Also, how do model parameters affect this transition time interval?
Given the observations of particle migration over a given time period as a point cloud data sampled at discrete time points, we use Topological Data Analysis, specifically persistent homology, to infer the transition time interval in which the particles undergo regime change. The topology of the particle configuration at any given time instance is captured by the persistent homology specifically \emph{Persistence Landscapes}. We localize (in time) when such a transition happens by conducting the statistical significance tests namely functional hypothesis tests on persistent homology outputs corresponding to subsets of the time evolution. This process is validated on a known collective behavior model of the self-propelled particles with the regime transitions triggered by changing the model parameters in time.  As an application, the developed technique was ultimately used to describe the transition in cellular movement from a disordered state to collective motion
when the environment was altered.

\end{abstract}
\nolinenumbers

\maketitle

\tableofcontents




\FloatBarrier
\section{Introduction}\label{sec:introduction}

In a system of many similar units such as flocks of birds, fish school, cells and molecules, the interactions between neighboring units leads to collective pattern of behaviour~\cite{VICSEK201271}.
The main question of interest is to detect the emergence of coherent state due to an environmental stimuli observed in such systems of interacting particles.
In this research we mainly refer to ``coherent state'' as the self-organization~\cite{SelforganizationFundamentCell} observed in biological system.
It is the emergence of an overall order in time and space of a given system that results from the collective interactions of its individual components. We want to address this problem that if the dynamical system is known only based on the observations of particle positions and velocities, then how to approach the questions of coherence seen in migration of particles due to environmental stimuli. The emerging field 
Topological data analysis (TDA) is a set of approaches that can help to understand complex data by studying its shape. Particularly, when there is a lack of information about what model mechanisms might be important, adopting a topological lens may be a useful approach for characterizing and comparing motion of biological groups ~\cite{ulmerTopologicalApproach2019}. The application of TDA has contributed to the understanding  not only for static data but also for time evolving data\cite{xianCapturingDynamicsTimeVarying2020} such as biological aggregation of insect swarm which vary from frame to frame in the movie of an experimental trial~\cite{ulmerTopologicalApproach2019}, detect differences in synchronization patterns in time-series output from networks of coupled oscillators~\cite{stolzPersistentHomologyTimedependent2017} etc.

Persistence homology is a tool from TDA that can allow comparisons of shape across the finitely sampled point clouds from a random process.
As the first step we present persistent homology as the technique
to uniquely visualize and identify the pattern changes of point cloud data sampled
from cell migration at discrete time steps even with noise in the data. We will review on the stable,  interpretable features of persistent homology, then provide  some illustrative examples in the form of simplicial complex, persistent homology descriptors corresponding to observed point cloud data sampled from ideal particle migration generated from a collective motion model at a temporal course.

The next step is to address the question how to detect whether and when the significant transition in pattern regimes due to an environment stimuli during an observed period of migration of system of particles, we propose a statistical approach in the context of functional topological summary called persistence landscape\cite{bubenikStatisticalTopologicalData}. A significant amount of work has been performed on application of integrating topological summaries with statistics~\cites{ Medina2016StatisticalMI, robinsonHypothesisTestingTopological2016, kovacev-nikolicUsingPersistentHomology2016}. Our approach of implementing a functional significance test with appropriate correction procedure on detecting the topological difference between snapshots of cell configuration is novel. Basically, at every discrete time step of the particle migration true shape-related features of particle configuration is unknown with only the availability of samples from different simulations.
Therefore we require statistical hypothesis testing method to decide if there is sufficient evidence to classify the shapes of the particle configuration at a certain time point is measurably different than the initial random particle configuration. So through the appropriate choice of significance test we can clearly quantify the “small” or “large” differences in shape features of such sampled particle configurations from a random particle configuration. The sequential implementation  of this test results in extracting useful insights on identifying the transition time of the given observed particle migration. Further it can be extended to the application of detecting pattern regime changes observed for cell movements in wound healing or cancer tumor.

The paper is organized as follows.
In \Cref{sec:problem-description}, first we will describe the synthetic data in which we model
the topology of random and coherent behaviour of particles.
 \Cref{sec:persistent-homology} will discuss the method to extract and visualize the topological features of a group of isolated points that represents a manifold as existing in persistent homology literature.
This is followed by \Cref{sec:hypothesis-pairwise-in-time} where we describe the proposed statistical testing framework with the construction of suitable time-varying
topological summary that can capture the critical transition in particle behaviour. The \Cref{sec:results_sim} will discuss the interpretation of results on the synthetic data and the summary of findings with experimental data. Furthermore, limitations, conclusion and future works will be discussed in \Cref{limitation} and \Cref{conclusion_future}.

\FloatBarrier
\section{Problem and data description}\label{sec:problem-description}
\FloatBarrier
\subsection{Problem statement}\label{sec:problem-statement}
One of the open question that arise in studying collective pattern behavior in biological system, either from the experimental or mathematical stand point is when particles move coherently, by which we mean they locally adjust their speed and direction to those of their neighbors in order to form complex aggregated patterns, what is the time interval on which coherence is achieved beginning from an incoherent state and how does the type of  external factors (experimental perspective) affect this time interval?

Let us consider a set of \(n\) particle trajectories obtained from the experimental observations or simulation of collective motion model such that each trajectory is represented by a sequence of time ordered locations given by \(X_{t}=\{\vec{r}_{i}(t)| i=\{1,2,\dots,K\}\}\subset \mathbb{R}^{2}\).
Here  \(\{X_{t}\}_{t\in m}\) can be considered as point cloud data in \(2D\) Euclidean space or high dimensional vector at \(t^{\text{th}}\) time unit  and \(m\) represents the discrete time units of the particle trajectory \(m=\{t_{1},\dots,t_{f}\}\).
Suppose there is a transition in particle configuration for a particular movement in which coherence is achieved at the final time point \(t_{f}\) beginning from a random state, then our goal is  to infer that transition time interval in which the particles undergoes the change based on given the observations \(X_{t}\)

\FloatBarrier
\subsection{Collective motion model}\label{sec:motion-model}

We develop the regime change technique and test it on the modified D'Orsogna particle model\cites{bhaskarTopologicalDataAnalysis2021,dorsognaSelfPropelledParticlesSoftCore2006}.
The model is a first-order discrete-time stochastic model describing the motion of ``cells'' (particles) in a two-dimensional rectangular domain with periodic boundary conditions under the influence of a stochastic propulsion force and a deterministic pairwise force.

\begin{equation} \label{eq:dorsogna-model}
\frac{\vec{r}_{i}[n+1] - \vec{r}_{i}[n]}{\Delta t} = 
\beta \hat{\vec{P}}_{i}[n]
+  
\alpha \sum_{j=1}^K \vec{F}( \vec{r}_j[n] - \vec{r}_i[n] )
\end{equation}
where \( \vec{r}_{i} = (x_i,y_i) \in [0,L] \times [0,L] \) are locations of \(K\) particles,
and the steps \(n = 0, \dots, N-1\) correspond to time instances \(t = n\Delta t\), where \(\Delta t = 0.02\).
The behavior of the model is controlled by the strength of the stochastic propulsion \(\beta \in [0.009,0.025]\) and the strength of pairwise interactions \(\alpha \in  [0.09,0.24]\).

The random unit-length vector \(\hat{\vec{P}}_{i}[n]\) models a run-and-tumble propulsion, i.e, constant-angle force (``run''), punctuated by an instantaneous change in direction (``tumble'') when the angle of \(\hat{ \vec{P} }\) is drawn uniformly at random from the interval \([0,2\pi]\).
The ``tumble'' steps occur regularly every \(N_T = \num{2500}\) time steps.
To avoid synchronized tumbles, each cell \(k\) is assigned a randomly chosen offset \(N_k\) so that its tumble steps occur at \(n = N_k, N_k + N_T, N_k + 2N_T,\dots\).

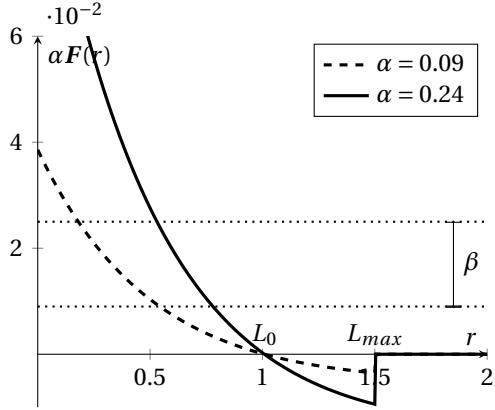
\begin{figure}[htb]
\begin{tikzpicture}[
  declare function={
    LA = 1/2;
    LR = 14;
    Lmax = 3/2;
    L0 = ln(4*LA/LR)*LA*LR/(LA-LR);
    func(\x,\L) = pow(e,-\x / \L)/ \L;
    P(\x) = ( -func(\x,LR) + func(\x,LA)/4 )*(\x <= Lmax);
  }
]
\begin{axis}[width=0.5\textwidth,
  axis x line=middle, axis y line=middle,
  ymin=-0.01, ymax=0.06, ylabel=\(\alpha \vec{F}(r)\),
  xmin=0, xmax=2, xlabel=\(r\),
  domain=0:2,samples=501, 
]

    \addplot [black,very thick,dashed] {0.09*P(x)};
    \addlegendentry{\(\alpha=\num{0.09}\)};
    \addplot [black,very thick,] {0.24*P(x)};
    \addlegendentry{\(\alpha=\num{0.24}\)};

    \addplot [black,thick,dotted] {0.009};
    \addplot [black,thick,dotted] {0.025};

    \node[black, anchor=south] at (axis cs:Lmax,0) {\(L_{max}\)};
    \node[black, anchor=south] at (axis cs:L0,0) {\(L_{0}\)};

    \node[black, anchor=west] at (axis cs:1.85,0.017) {\(\beta\)};

    \draw[draw = black, |-| ] (axis cs:1.85,0.009) edge (axis cs:1.85,0.025);

\end{axis}
\end{tikzpicture} 
\caption{Magnitude of the pairwise interaction force~\eqref{eq:pairwise-force}, depending on the distance between particles \(r\) for the extents of the considered range of interaction strength \(\num{0.09} \leq \alpha \leq \num{0.24}\).
Force of a positive sign is repelling. For context, horizontal lines mark the extents of the considered range of propulsion strength \(\num{0.9e-2} \leq \beta \leq \num{2.5e-2}\).}\label{fig:pairwise-force}
\end{figure}

The pairwise interaction function models ``soft core'' particles with ``hair'', which move freely at longer distances, but ``stick'' when their ``hair'' comes into contact. 
If pushed too close, where their ``core'' touches, they repel.
This interaction is modeled by the function
\begin{equation}
\vec{F}(\vec{v}) = \hat{\vec{v}}\left[
-\frac{1}{L_{A}} e^{-\norm{\vec{v}}/L_{A}} + 
  \frac{1}{4 L_{R}} e^{-\norm{\vec{v}}/L_{R}} 
  \right]
  \text{ if } \norm{\vec{v}} \leq L_{max}, 
\label{eq:pairwise-force},\end{equation}
and \(\vec{F}(\vec{v}) = \vec{0}\) otherwise, where \(\hat{\vec{v}} = \vec{v}/\norm{\vec{v}}\) is the unit-vector in the direction of the argument.
The resulting force vector \(\vec{F}(\vec{r}_j - \vec{r}_i)\) repels particles when they are nearer than a critical distance (``core''), i.e.,
\begin{equation}
\norm{\vec{r}_j - \vec{r}_i} < L_0 \coloneqq \frac{L_A L_R}{L_A - L_R}\ln\frac{4L_A}{L_R},
\end{equation}
and attracts them (``stickiness'') if their distance is between \(L_0\) and \(L_{max}\).
In all presented work we set \(L_{max} = 1.5\),\(L_A = 14\), and \(L_R = 1/2\) resulting in \(L_0 = 1.009\).
As seen in~\Cref{fig:pairwise-force}, the cutoff \(L_{max}\) is chosen so that the force is significantly discontinuous at that distance, mimicking the finite ``hair'' length.

\begin{figure}[htb]
\centering\includegraphics[width=.8\textwidth]{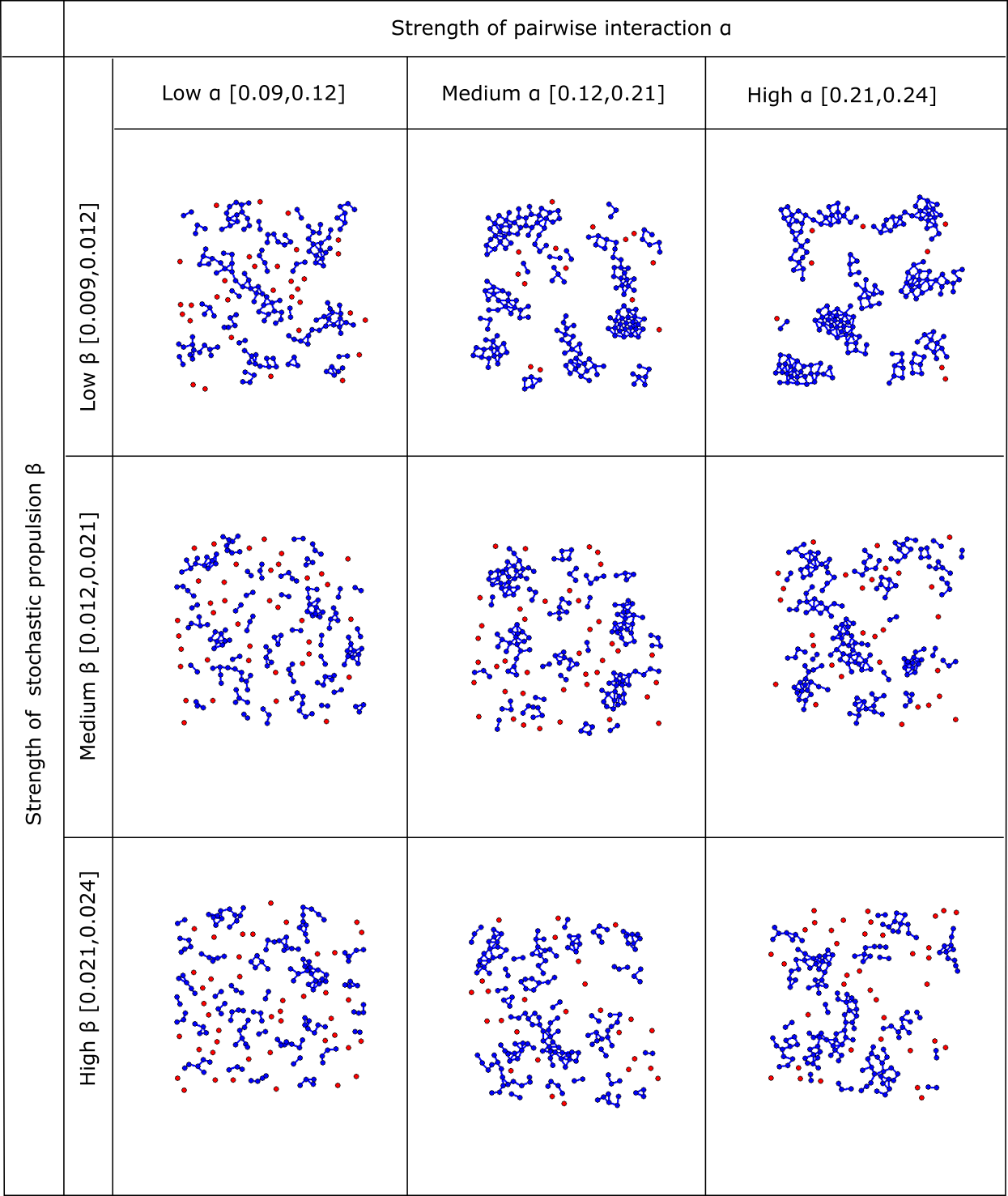}
\caption{Examples of configurations of  particles in a square, periodic box with sides \(L=20\), evolved from a uniformly random states for \(N=50000\)  steps under the modified D'Orsogna model~\eqref{eq:dorsogna-model}. 
Columns correspond to the change in strength of the pairwise interaction \(\alpha\), rows to the change in strength of the self-propulsion \(\beta\).
At the final time step, particle having interaction (\(\vec{F}(\vec{r}_j - \vec{r}_i)\)) with neighbors  are shown in blue and particles with no interaction (\(\vec{F}(\vec{r}_j - \vec{r}_i)=0 \)) are shown in red. }
\label{fig:particle-regimes}
\end{figure}

By varying the parameters \(\alpha\) and \(\beta\) the particle cloud evolves toward three different types of spatial configuration, shown in~\Cref{fig:particle-regimes}:
\begin{itemize}
\item clustered, for \(\beta\in[0.009,0.012]\) and  \(\alpha\in[0.21,0.24]\),
\item disordered, for \(\beta \in[0.021, 0.024]\) and  \(\alpha \in [0.09,0.12]\), and
\item so-called branched-clustered otherwise.
\end{itemize}
The ranges above are taken as approximate, since there is no purely-objective way to define the difference between the regimes, but they will suffice for the work we present.

\FloatBarrier
\section{Persistent homology of point clouds}\label{sec:persistent-homology}

Let us consider a set of \(P\) particle trajectories obtained from the dynamical system in \Cref{eq:dorsogna-model}.
At any time \(t\) the points from the trajectories form a point clouds (snapshot) \(X_{t}=\{\vec{r}_{i}(t)\}_{i=1}^P\subset \mathbb{R}^{2}\).
We process this data as a set of \emph{snapshots}, treating each point cloud \(X_{t}\) separately.

The systematic way of forming shapes from the point cloud data can be done through forming simplices.
If we assume that \(d\) points in \(\mathbb{R}^2\) are linearly independent (with respect to any choice of the origin), the geometric \(d\)-dimensional simplex is the smallest convex subset of \(\mathbb{R}^{2}\) containing those points.
For example, \(0\)-simplex is a point, \(1\)-simplex is a line, \(2\)-simplex is triangle.
More abstractly, a \(d\)-simplex is an ordered set (tuple) of points.

A collection of simplices, along with a description of how they attach to each other in the case of abstract simplices, is called the simplicial complex.
For example, a tetrahedron can be thought of as a simplicial complex because it describes how vertices (\(0\)-simplices) attach to each other via edges (\(1\)-simplices), and how faces (\(2\)-simplices) attach to the edges.

Homology is an algebraic process that associates a collection of vector spaces to a topological space, such as a simplicial complex, in such a way that maps between those vector spaces mirror the connections between simplices.
In this way, topological properties such as homology groups, which count the number of connected components (\(H_0\) group) or the number of holes (\(H_1\) group) in the complex, can be represented by algebraic properties, such as dimensions of image and null spaces of the maps in homology, and computed using linear algebra.
The linear-algebraic algorithms for storing simplicial complexes and computing their homology vary in implementation, efficiency, and complexity, but there are many freely-available packages that allow for an easy application.

The process of converting geometric data to a (family of) simplicial complexes is called \emph{filtration}, resulting in a \emph{filtered} simplicial complex.
Starting from a set of points \(\vec{r}_k \in \mathbb{R}^2\), \(k=1,\dots,K\) and a scale parameter \(\varepsilon\), the filtration algorithm is used to decide what \(d\)-simplices to associate with subsets of \(d+1\)-points (individual points are always \(0\)-simplices).
For example, the \emph{Vietoris--Rips} (VR) filtration ~\cites{ulmerTopologicalApproach2019, ghristBarcodesPersistentTopology2008} assigns \(1\)-simplices (edges) whenever \(\norm{\vec{r}_i - \vec{r}_j} \leq \epsilon\), \(2\)-simplices (faces) whenever a triplet of \(1\)-simplices forms a triangle (shares endpoints), and so on.
Alternative filtrations (\v{C}ech, witness)~\cites{edelsbrunner2008, Edelsbrunner2010} share many properties, but can differ in the amount of theoretical foundation for their use and in their computational complexity. 
Computational algorithms typically form the filtration across a predetermined range of scales \(0 \leq \epsilon \leq \epsilon_{max}\) and, sometimes, with a predetermined resolution, resulting in a vector of \(\epsilon_i\) values for which the simplicial complexes (and their homologies) are available.
For our purposes, the described VR filtration suffices and is used in the rest of this work.

\begin{figure}[htb]
\begin{center}
\includegraphics[width=.9\textwidth]{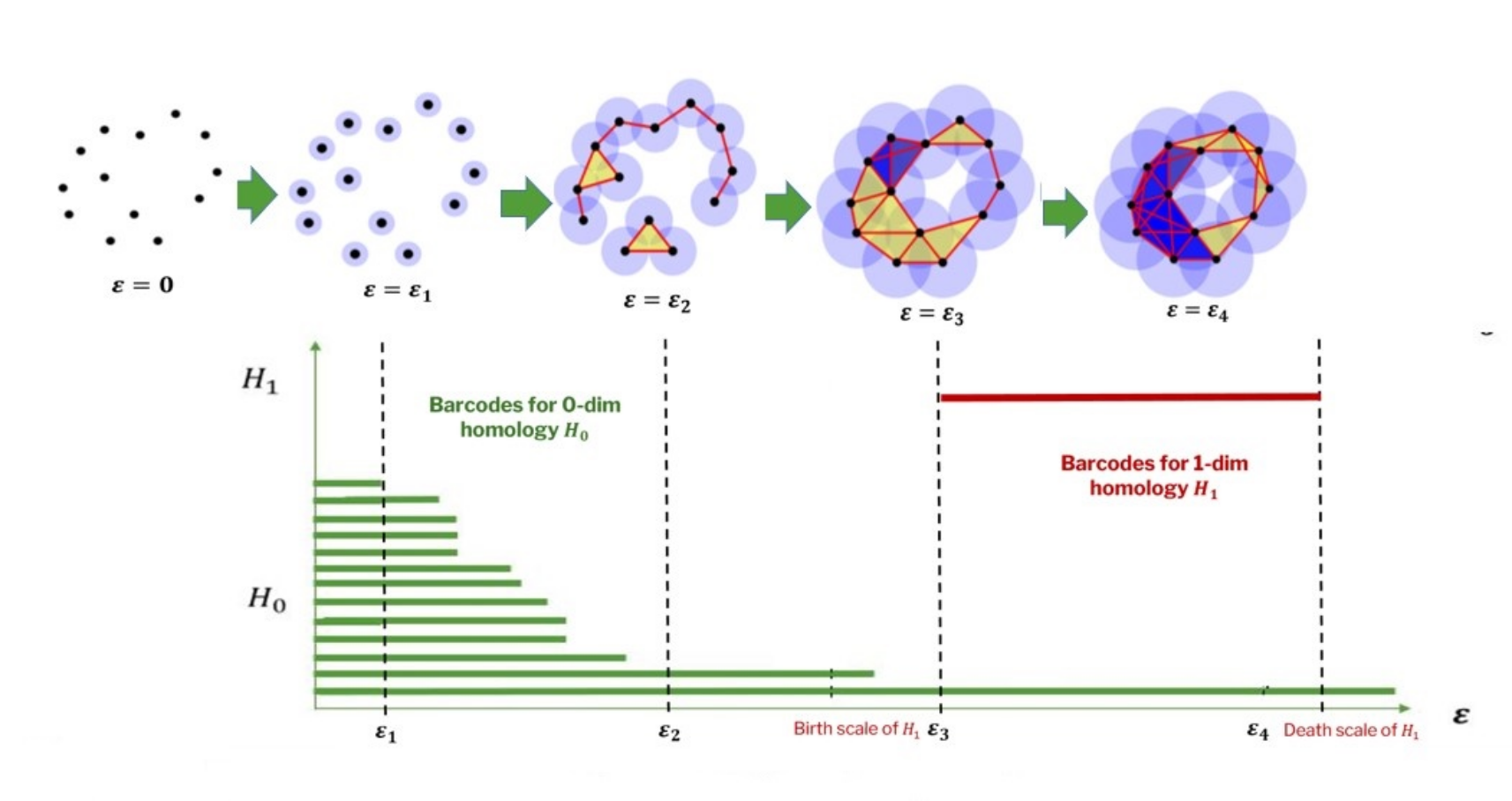}
\end{center}
\caption{Barcode representation Persistent homology for a point cloud using the VR complex}
\label{fig:barcode-illustration}
\end{figure}

The \emph{persistent homology} explains how \(d\)-dimensional ``holes'' appear and disappear in the point cloud as the parameter \(\epsilon\) is varied.
For example, \(H_{0}\) group corresponds to connected components in the simplicial complex.
When \(\epsilon \approx 0\) it is smaller than the distance between any two points; therefore there are no edges or higher simplices, and each point is counted as a separate component.
As \(\epsilon\) increases, those points closes to each other will start connecting via edges, so the number of connected components will reduce.
Depending on the arrangement of points, some subset of points may end up circularly connected, without a face being inscribed between them.
In this way a \(H_{1}\) element, or a loop, appears (is \emph{born}); it disappears (\emph{dies}) when the \(\epsilon\) parameter is large enough so that additional edges form that ``bridge the gap'' across the loop.
An example of this process is shown in the top panel of \Cref{fig:barcode-illustration}.

Changes in homology groups with respect to \(\varepsilon\) can be represented in a large variety of ways~\cites{adamsTopologyAppliedMachine2021, edelsbrunnerTopologicalPersistenceSimplification2002, edelsbrunner2008}.
Barcodes and persistence diagrams are most common visual representations, however, additional representations such as persistence landscapes and persistence images have their use if the end goal is not visualization but rather statistical analysis or machine learning~\cites{mileykoProbabilityMeasuresSpace2011, bukkuriApplicationsTopologicalData2021}.

As seen \Cref{fig:barcode-illustration}, the \(1\)-dimensional homology class \((H_{1})\) born at \(\varepsilon_{3}\) (birth scale) and dies at \(\varepsilon_{4} \) (death scale) can be represented as a barcode of length  \(\varepsilon_{4}-\varepsilon_{3}\) which is called persistence length of \(H_{1}\) feature.
These barcodes can also be represented as a \emph{persistence diagram}, which is a scatter plot of birth--death pairs, with each point corresponding to one feature in a particular \(H_{d}\) barcode. 
The distance between the point and the diagonal line is the \emph{persistence}; points further from the diagonal are interpreted as corresponding to features that are robust with respect to small perturbations of the input data, while those close to the diagonal are sometimes referred to as \emph{topological noise}, since they appear and disappear very quickly.

To compare topology of two different point clouds, we need to define a distance-like function between two representations of persistence homology.
While it is possible to define such a function between persistence diagrams~\cites{mileykoProbabilityMeasuresSpace2011,oudotPersistenceTheoryQuiver2015,bubenikStatisticalTopologicalData}, representing the diagrams by a persistence landscape, can use a wider range of already-developed techniques for data mining and statistical analysis of (continuous) real-valued functions.

\begin{figure}[htb]
  \centering\includegraphics[width=1\textwidth]{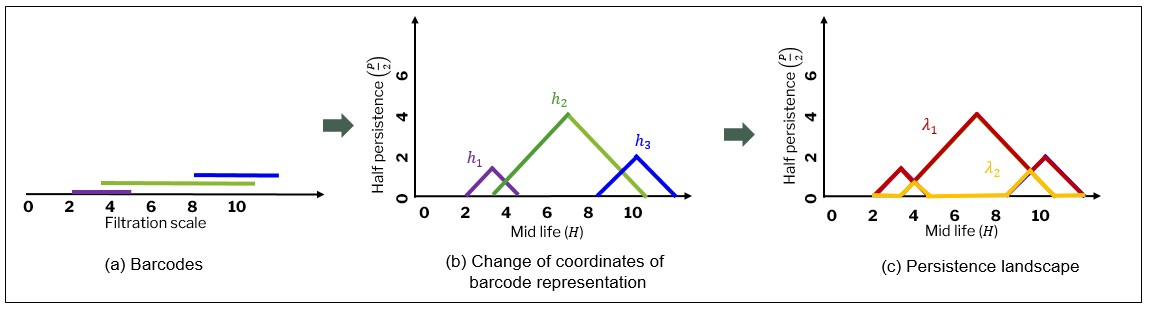}
  \caption{ Construction of persistence landscape from barcode corresponding to a point cloud data.  }
  \label{fig:PL_construction}
  \end{figure}

The persistence landscape is defined to each birth-death pair in the persistence diagram \(\{ (b_i,d_i) \}_{i=1}^M\) we associate the persistence function \(h_{(b,d)}:\mathbb{R} \to [0,\infty)\)
\begin{equation} \label{eq:pers-func} 
 h_{(b,d)}(\epsilon) \coloneqq \max\lbrace P/2 - \abs{ \epsilon - H }, 0\rbrace 
\end{equation} 
where \(P = d - b\) is called \emph{persistence}, and \(H = (b+d)/2\) is called \emph{mid$-$life} as shown \Cref{fig:PL_construction}

The persistence function \(h_{(b,d)}\) is zero for \( \epsilon \not = [b,d]\), inside the interval rises with a unit slope to \(P/2\), and then falls back to zero.
To avoid subindices, we re-label these functions as \(h_i \equiv h_{(b_i,d_i)}\).

The \emph{persistence landscape} (PL) is the collection of the upper envelopes of the set of persistence functions \(h_i\).
More precisely, the \(m\)-th upper envelope of the set of curves \(\{ h_i(\epsilon) \}_{i=1}^M\) is defined pointwise as
\begin{equation}
  \lambda_m(\epsilon) \equiv \lambda(\epsilon, m) \coloneqq \left\lbrack \{ h_i(\epsilon) \}_{i=1}^M \right\rbrack_{(m)},
\end{equation}
where notation \([S]_{(m)}\) for \(m \leq M\) stands for the \(m\)-th largest value in the set \(S\) of \(M\) elements (the \(m\)-th order statistic).  

Finally, the \emph{persistence landscape contour} (PLC) is the mean of the first \(M' \leq M\) envelopes in the landscape
\begin{equation}
  L(\varepsilon) \coloneqq \frac{1}{M'}\sum_{m=1}^{M'}\lambda_{m}(\epsilon).
\end{equation}

As continuous functions on a (practically) bounded interval, PLC can be endowed with usual \(\norm{\cdot}_p\)-norms which can then be used to measure the difference between two data sets~\cites{bubenikStatisticalTopologicalData,wadhwaTDAstatsPipelineComputing2018}.
Note that there are data sets that are different from each other (including in the number of points) that may produce identical persistence diagrams, so the zero-distance between two PLCs may not imply that the original data sets are the same, although if all subsets of data produce identical PDs then the data must be identical~\cite{robinsonHypothesisTestingTopological2016}.

Even though the sequence of transformations from PD to PLC is substantial, the PLC is statistically stable which enables its use as an input in statistical analysis.
It is shown in ~\cites{bubenikStatisticalTopologicalData,bubenikPersistenceLandscapesToolbox2017} that \emph{persistence landscapes} \((\lambda(\epsilon,m))\) are Borel random variables with values in separable Banach space \(L^{p}(\mathbb{N} \times \mathbb{R})\) for \(p \geq 1\) and for any continuous linear functional \(f \in L^{q}\) such that \(\frac{1}{p} + \frac{1} {q} =1\), the random variable \(f(\lambda(\epsilon,m))=\sum_{m=1}^{M} \int_{\mathbb{R}} \lambda(\epsilon,m) d\epsilon\) satisfies the statistical properties such as stability/continuity (small perturbation in the point cloud results in small perturbation in the PL distance), and consequently convergence, strong law of large numbers (SSLN), and central limit theorem (CLT). 
The averaging process that results in PLCs is an example of such a functional that is explicitly analyzed in~\cite{bubenikStatisticalTopologicalData}, along with the weighted-average variants.

As mentioned, the \emph{topological noise} are features with the low persistence, which appear close to the diagonal in the PD.
Noise rejection of such features can be done in (at least) two ways.
First, we can remove the points with persistence smaller than some noise-rejection threshold.
Many implementations of PH offer this option as a flag in the computational algorithms used to compute PH\@.
In regions of \(\epsilon\) where only topological noise is present, this strategy results in PLC being equal to zero.
The second approach is to choose lower \(M'\) in the averaging step of PLC\@. 
This tunes the noise rejection locally.
For scales \(\epsilon\) where there are many very persistent features, topological noise does not contribute to the contour.
For \(\epsilon\)s with only low-persistence features, the mean will reflect their presence, instead of being identically zero.

\FloatBarrier
\section{Inferring regime transition for multiple observed particle movements}
\label{sec:hypothesis-pairwise-in-time}
In this section we describe how statistical significance tests namely functional hypothesis tests can be applied to the resulting time varying persistent homology outputs in order to detect whether model parameters affect the pattern changes in the motion and estimate time points at which changes occur. We repeatedly apply hypothesis tests on PLCs at each filtration scale $\epsilon$ corresponding to pairs of snapshots in order to identify whether the ``topological distributions'' of particles in two snapshots are similar or not. A repeated application of a pairwise hypothesis test may result in inflated confidence about (non-)rejection of the particular null-hypothesis.
Therefore, we first describe the hypothesis test used without regard for its repeated use, and then describe the strategy for the repeated tests that takes into the account the inflation in the confidence values.

\FloatBarrier
\subsection{Pointwise hypothesis test for comparing two static point clouds}
\label{sec:pointwise-two-time-test}

Each simulation (or experiment) can be thought of as a time-continuous sample from a hypothetical population of such simulations.
We model our data collection as two (non-commutative) operations: first, retrieving \(S\) samples of time-continuous populations, and second, retaining only two snapshots, \(A\) and \(B\) representing two model parameters, from each time-continuous sample.
This gives rise to a collection of point clouds, \(X_{A,s}\), \(X_{B,s}\) for \(s=1,2,\dots, S\).
We then compute the PLCs) for each of the samples, resulting in two collections of \(\epsilon\)-continuous curves \(L_{A,s}(\epsilon)\) and \(L_{B,s}(\epsilon)\), with a common domain \(0 \leq \epsilon \leq \epsilon_{max}\) at each time step of the simulation.
Features that persist for \(\epsilon > \epsilon_{max}\) will be interpreted as persisting indefinitely.
In our problem setting, we are interested to check whether PLC functions \(A\) and \(B\) differ significantly at each time and thereby infer the existence of effect due to parameter change as well as the time points at which regime change occurs.

Functional Data Analysis (FDA) is the branch of statistical inference concerned with data comprising curves (rather than scalars or vectors).
For example, the pointwise t-test~\cite{Ramsay} is a parametric hypothesis test that assesses whether two groups of curves that are sampled from pointwise-normal populations have the same mean curve. Pointwise, we cannot claim that PLCs are distributed normally, or that their differences are distributed normally. Non-parametric rank-based univariate two-sample comparison procedure is relatively powerful and commonly used when there is no assumptions made about the distributions of data within each sample~\cites{Wilcoxon,NonparametricFDA,Sign_Wilcoxon_and_Mann-Whitney}. Pointwise with respect to \(\epsilon\), we use the non-parametric Wilcoxon two-sample rank test~\cite[\S 3.1]{Hollander2013} that is applicable to our data, which satisfies the following assumptions:

\begin{itemize}
    \item Simulations are independent random events, as they were generated from random initial configurations
\item Data points are $\epsilon-$ continuous curves (PLC) corresponding to two groups of simulations which are simulated from two different sets of parameter
\item Each group of PLCs is independent from  each other as they corresponds to PLCs computed based on simulations of two different sets of parameters and they are continuous populations
\end{itemize}

 Suppose that for any fixed \(\epsilon\), the distribution functions corresponding to \(L_{A}(\epsilon)\) (population 1) and \(L_{B}(\epsilon) \) (population 2)  be \(F_{\epsilon}^{A}\) and \(F_{\epsilon}^{B}\)
respectively. 

A \emph{two-sided test} distinguishes between the following  \(\epsilon\)-pointwise hypotheses
\begin{equation}\label{eq:null-pointwise} 
    H_{0}(\epsilon):\, F_{\epsilon}^{A}=  F_{\epsilon}^{B} \quad \text{vs. }\quad H_{1}(\epsilon):\,  F_{\epsilon}^{A} \neq F_{\epsilon}^{B},
\end{equation}
with \(H_{0}(\epsilon)\) asserts that the \(L_{A}(\epsilon)\) variable and the \(L_{B}(\epsilon)\) variable have the same probability distribution, but the common distribution is not specified. The \(H_{1}(\epsilon)\) asserts that \(L_{A}(\epsilon)\) tends to be larger (or smaller) than \(L_{B}(\epsilon)\) \cite{Hollander2013}. In other words, null hypothesis corresponds to no change in the PLC distributions between two snapshots at the scale \(\epsilon\)  and alternative hypothesis corresponds to the presence of the change at the scale \(\epsilon\).

To compute the Wilcoxon two-sample rank sum test statistic, we use the ascending rank function
\begin{equation}
\rank(x_s) = r
\end{equation}
that assigns integer ranks \(r = 1,\dots,S, S+1, \dots, 2S\) to order the combined samples of \(L_{A}(\epsilon)\) and \(L_{B}(\epsilon)\) from smallest to largest, i.e., \(\rank(x_1) < \rank(x_2)\) if \(x_1 < x_2\).
We also use the positive part function
\begin{equation}
\pos{x} \coloneqq \max\{x, 0\}.
\end{equation}

For each \(\epsilon\) the the Wilcoxon signed rank test statistic \(T(\epsilon)\) will be the summation of ranks denoting the rank of \(L_{B}(\epsilon)\) (population 2) in the combined sample of size $2S$
    \begin{equation}
    T(\epsilon) \coloneqq \sum_{s=S}^{2S} \pos{ L_{B}(\epsilon)} \rank  \abs{ L_{B}(\epsilon)}   = \sum_{\substack{s=S \\  L_{s}(\epsilon) 
 >0 }}^{2S} \rank( \abs{ L_{B}(\epsilon)} ),
\end{equation}

At the level of \(\alpha\) significance, we can \emph{reject} the null hypothesis \(H_{0}\)  if 
\begin{equation}
T(\epsilon) \geq w_{\alpha/2} \quad \text{or} \quad T(\epsilon) \leq S(2S+1)/2 - w_{\alpha/2}
\end{equation}
with probability of \(\frac{\alpha}{2}\) in
each tail of the null distribution of \(T(\epsilon)\).

The critical value \(w_{\alpha/2}\) is chosen to make the type I error probability equal to \(\frac{\alpha}{2}\).
The test is available in R language using \texttt{wilcox.test} command, and the choice of critical value using \texttt{psignrank}.

Furthermore, for sample sizes greater than 25, a normal approximation the Wilcoxon signed-rank distribution is valid~\cite{wackerly02} and therefore one can use instead the standardized test statistic
\begin{equation}
\hat{T}(\epsilon)=\frac{T(\epsilon)-\E[T(\epsilon)]}{\sqrt{V(T)}}=\frac{T(\epsilon)-S(2S+1)/2}{\sqrt{S^{2}(2S+1)/12}},\label{eq:signed-rank-test-parametric}
\end{equation}
which asymptotically follows a \(N(0,1)\) normal distribution.
Therefore, we can reject \(H_{0}\) if 
\begin{equation}
    \abs{\hat{T}(\epsilon)}\geq z_{\alpha/2},
\end{equation}
where the critical value is computed using the normal distribution.


Both \(T(\epsilon)\) and \(\hat{T}(\epsilon)\) are test statistics computed pointwise for the purpose of comparing the two sample groups \(L_{A,s}(\epsilon)\) and \(L_{B,s}(\epsilon)\) at a specific scale.
Typically, one is interested in evaluating the difference between groups \(A\) and \(B\) across \emph{all} \(\epsilon\) scales.

The most straightforward way of addressing this demand is to select a single scale for which to perform the test. We define the global test statistics as
\begin{align}
\hat T &\coloneqq \max_\epsilon \abs{ \hat{T}(\epsilon) }
\end{align}
Such choice appears to correspond to the global versions of null hypotheses~\eqref{eq:null-pointwise}
\begin{equation}\label{eq:null-global} 
    H_{0}:\, \forall \epsilon,\ F^{A}(\epsilon)= F^{B}(\epsilon) \quad \text{vs. } \quad H_{1}:\, \exists \epsilon, \ F^{A}(\epsilon)\neq  F^{B}(\epsilon).
\end{equation}


Computationally, this amounts to choosing a grid of values \(\epsilon_k\), \(k = 1,\dots, K\),  from \([0,\epsilon_{max}]\) and computing the pointwise test statistics, and rejecting the global null hypothesis~\eqref{eq:null-global} if at least one pointwise hypothesis~\eqref{eq:null-pointwise} is rejected.
Ideally, increasing the resolution of the grid \(K\to\infty\) would result in a more reliable testing procedure.
At the same time, simultaneously testing \(K \to \infty\) different hypotheses between the samples of \(L_{A,s}\) and \(L_{B,s}\) at each filtration scale $\epsilon$ increases the chance of committing a False-Positive error. To overcome this multiplicity problem due to correlation, we use the Westfall-Young correction method introduced by Cox and Lee \cite{Westfall}.

We assume that all PLC curves of two groups \(L_{A,s}\) and \(L_{B,s}\)  have a common set of evaluation points $\epsilon$, $ 0 \leq \epsilon \leq \epsilon_{max}$. The multiple comparison procedure that controls for family-wise error rate is applied on the p-value $p(\epsilon)$ obtained from pointwise uni variate Wilcoxon two-sample rank sum test corresponding to the null hypothesis $H_{0}(\epsilon)$. To define the family-wise error rate, consider
\begin{align}
C_{m}=\{ \epsilon :\;\; H_{0}(\epsilon)\;\; \text{is true, }\;\; 0 \leq \epsilon \leq \epsilon_{m}\}
\end{align}

which is the set of $\epsilon$ values for which $H_{0}(\epsilon)$ is true. Then, the family wise error is given by the probability:
\texttt{pr\{reject $H_{0}(\epsilon)$ for any $\epsilon$ in $C_{m}$\}}

We want to make the family-wise error rate less than or equal
to a nominal significance level $\alpha$ regardless of what is the set $C_{m}$ of true null hypotheses. We have provided the summarised steps for the procedure to control the family wise error rate which is the probability of committing at least one False-Positive error to be less than or equal to $\alpha$ as follows:\cite{Westfall}, \cite{Keser2014ComparingTM}

\begin{enumerate}
    \item Obtain the unadjusted \emph{p-}values by performing the pointwise Wilcoxon tests at every  $\epsilon$, $ 0 \leq \epsilon \leq \epsilon_{max}$ on the given data. The unadjusted p-values are ordered from min to max $p_{\epsilon_{1}}<p_{\epsilon_{2}}, \cdots p_{\epsilon_{k}}$
    \item In order to randomize the data and obtain adjusted p-values\emph{p-} values, start the counting variables ($R_{i}=0, i=1,2, \cdots, k$. Then \emph{p}-values are computed from a randomized data set which are given by \emph{$p^{*}$}. These values are kept in the same order as the sorted \emph{p-} values for the given data in step 1.
    \item Then define the successive minima as follows and update $R_{i} \longleftarrow R_{i+1}$ if $q_{i}^{*} \leq p_{i}$ 
    
    \begin{align*}
        q_{k}^{*}=p_{\epsilon_k}^{*}\\
        q_{k-1}^{*}=min(q_{k}^{*}, p_{\epsilon_{k-1}}^{*})\\
        \vdots\\
 q_{1}^{*}=min(q_{2}^{*}, p_{\epsilon_{1}}^{*})\\
    \end{align*}
 \item After repeating the randomization shown in Step 2 and 3 $N$ times ($N \approx 1000$), the adjusted \emph{p-}value is the proportion of $q_{i}^{*} \leq p_{i} $computed as $\hat{p}_{i}^{N}=R_{i}/N$ along with an additional constraint which apply monotonicity using successive maximization as follows:
    \begin{align*}
        \hat{p}_{1}^{N}=\hat{p}_{1}^{N}\\
        \hat{p}_{2}^{N}=max(\hat{p}_{1}^{N}, \hat{p}_{2}^{N})\\
        \vdots\\
 \hat{p}_{k}^{N}=max(\hat{p}_{k-1}^{N}, \hat{p}_{k}^{N})\\
    \end{align*}
 
 After applying monotonicity, the simulation bases $p_{j}^{N}$ reasonably approximate the actual values of $\hat{p_{j}}$ for sufficiently large $N (\approx 10000)$ 
    
\end{enumerate}

\section{Results and discussion on application to the simulation and experimental data}
\label{sec:results_sim}

\begin{figure}[htb]
\begin{center}
\includegraphics[width=.7\textwidth]{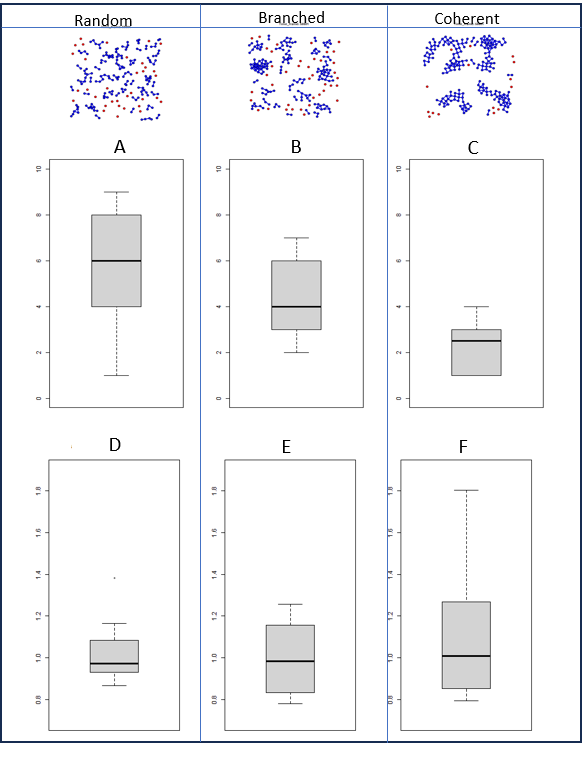}
\end{center}
\caption{Box plots for average number of overlapping landscapes (A-C) and persistence length of features (D-F) corresponding to particle migration namely random, branched, coherent with parameters $(\alpha= 9 \times 10^{-2}$, $\beta =2.1 \times 10^{-2}$ ),($\alpha=2.1 \times 10^{-1}$,  $\beta =2.1 \times 10^{-2}$), ($\alpha=2.4 \times 10^{-1}$,  $\beta =9 \times 10^{-3}$)  respectively.}
\label{fig:PL-sim-illustration}
\end{figure}

As the first step we present persistent homology as the technique
to uniquely visualize and identify the different pattern changes of a snapshot in the particle migration generated based on \Cref{eq:dorsogna-model}.  Based on the quantification of number of overlapping persistence landscape and the persistence length of three different simulations namely random, branched and coherent we can observe the following important observations \Cref{fig:PL-sim-illustration}:
\begin{itemize}
\item the average number of overlapping landscapes of random > branched >coherent configurations implying the existence of one-dimensional holes decreases.

\item the average half-persistence length (or altitude of persistence landscape) of largest feature for random < branched < coherent
\end{itemize}
These two observations based on persistence landscape implies, the existence of several short-lived features for random configuration
the existence of long lived ( high persistence length) features for coherent configuration. Therefore, this persistent homology representation can provide an insight on distinguishing multivariate shape feature changes that can be seen in particle migration at varied parameter changes (analogously distributional changes in cell migration due to varied environmental factors)

\subsection{Detect the effect due to interaction parameter changes in the particle migration}

\begin{figure}
  \centering\includegraphics[width=1.0\textwidth]{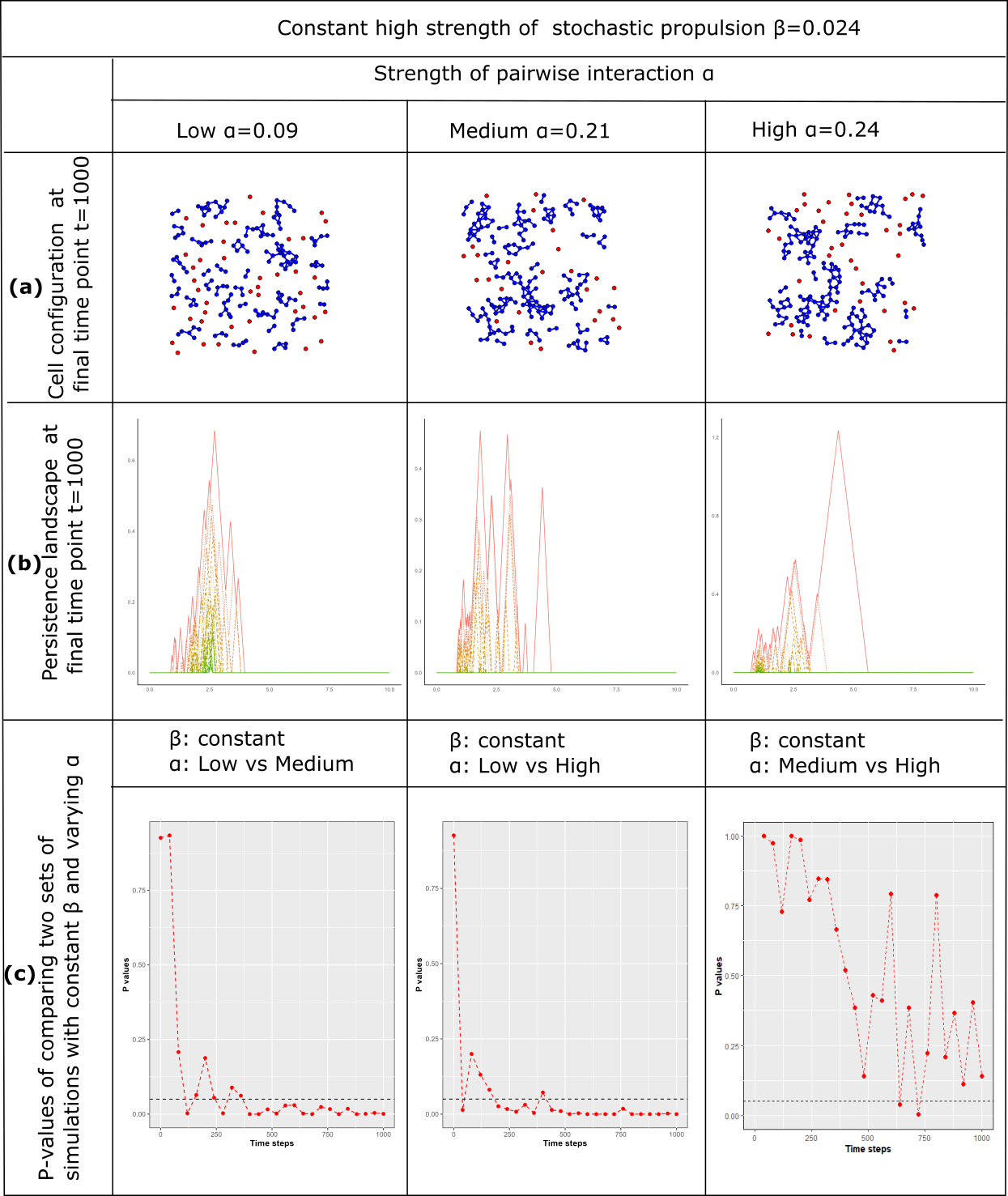}
\caption{(a) One of 100 samples of simulations generated with constant high $\beta\,(2.4 \times 10^{-2}) $ parameter and varying $\alpha $ parameters as \textbf{Low},  \textbf{Medium}, \textbf{High} signifying a final configurations of random, branched-cluster, branched-cluster phases respectively. (b) the persistence landscapes corresponding to the respective final configurations. (c) The adjusted global p-values based at consecutive time points implemented when pairwise comparing the samples of simulations due to the effect of varying $\alpha $ parameters as \textbf{Low},  \textbf{Medium} and \textbf{High}}
\label{fig:const_high_beta}
\end{figure}

We can examine whether different time steps have significantly differ in particle configuration due to the effect of varying strength of stochastic propulsion and determine whether or not a regime transition exist through implementing the non-parametric functional hypothesis test approach as developed in \Cref{sec:pointwise-two-time-test}. It can be determined from \Cref{fig:const_high_beta} that for \textbf{Low} vs \textbf{Medium}, \textbf{Low} vs \textbf{High}, the \emph{p-}values $<0.05$ consistently after $t=500$ which implies when there is high randomness in the particle configuration, the effect due to increase in strength of pairwise interaction among particles is evident but over the similar time interval. The \textbf{Medium} and \textbf{High} effect corresponds to a branched final configuration, the p-values throughout the whole time domain indicates there is no significant difference between these two movements and at each time the configuration is topologically similar.

\subsection{Detect the effect due to propulsion parameter changes in the particle migration}

\begin{figure}
  \centering\includegraphics[width=1.0\textwidth]{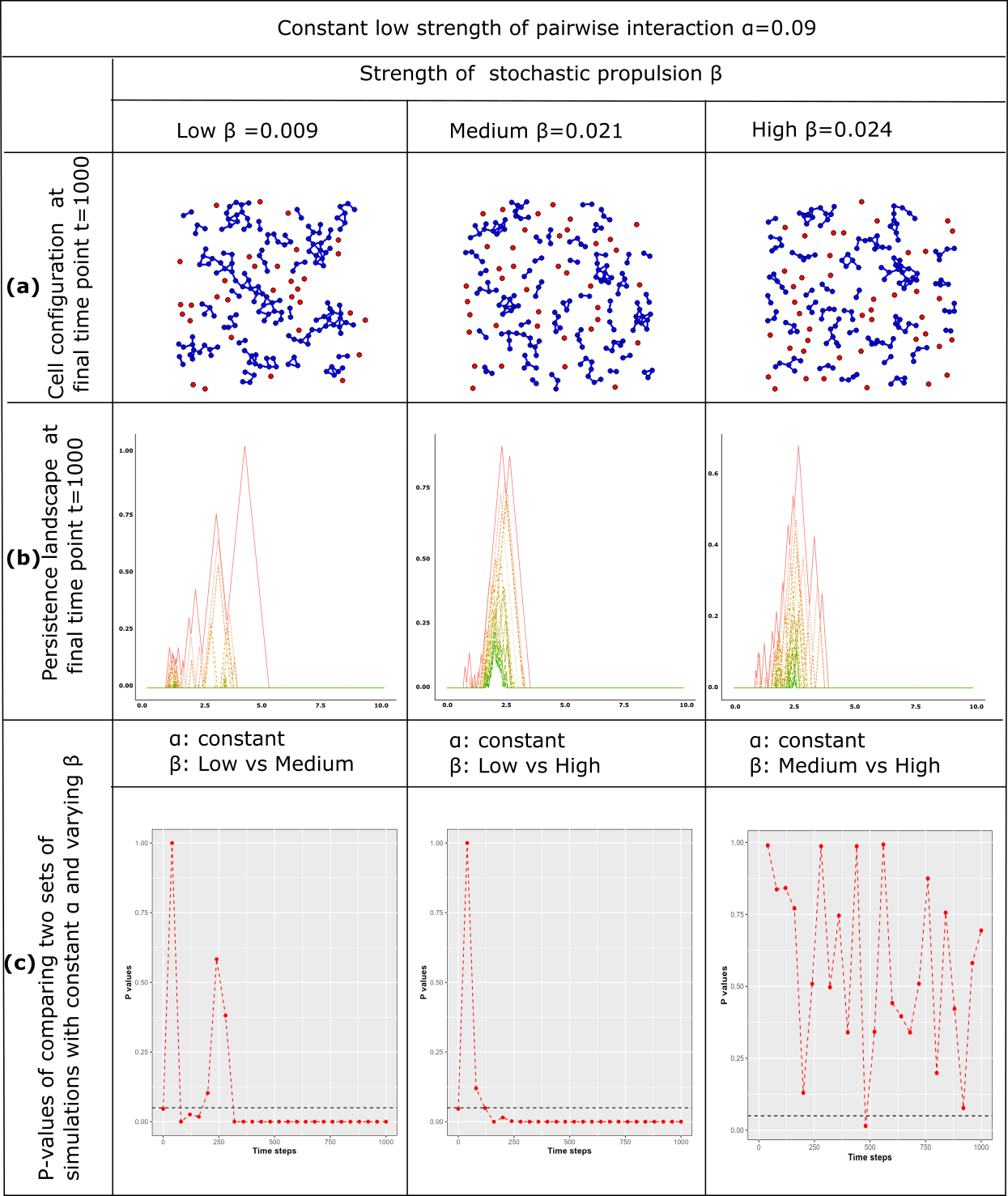}
\caption{(a) One of 100 samples of simulations generated with constant low $\alpha\,(9 \times 10^{-2}) $ parameter and varying $\beta $ parameters as \textbf{Low},  \textbf{Medium}, \textbf{High} signifying a final configurations of random, branched-cluster, branched-cluster phases respectively. (b) the persistence landscapes corresponding to the respective final configurations. (c) The adjusted global p-values based at consecutive time points implemented when pairwise comparing the samples of simulations due to the effect of varying $\alpha $ parameters as \textbf{Low},  \textbf{Medium} and \textbf{High}}
\label{fig:const_low_alpha}
\end{figure}

It can be determined from \Cref{fig:const_low_alpha} that for \textbf{Low} vs \textbf{Medium}, \textbf{Low} vs \textbf{High} the \emph{p-}values $<0.05$ consistently in the time intervals $[300,1000]$ and $[200,1000]$ which implies when there is low strength of interaction in the particle configuration, the effect due to increase in strength of propulsion among particles is evident. When comparing the \textbf{Medium} vs \textbf{High} randomness, there is no significant effect due to the change of strength of propulsion at any time step. The test precisely detect no significant changes between two complete random movements.

\FloatBarrier
\subsection{Practical implications of the study}
\begin{figure}
\centering
\begin{subfigure}{0.15\textwidth}
    \includegraphics[width=\textwidth]{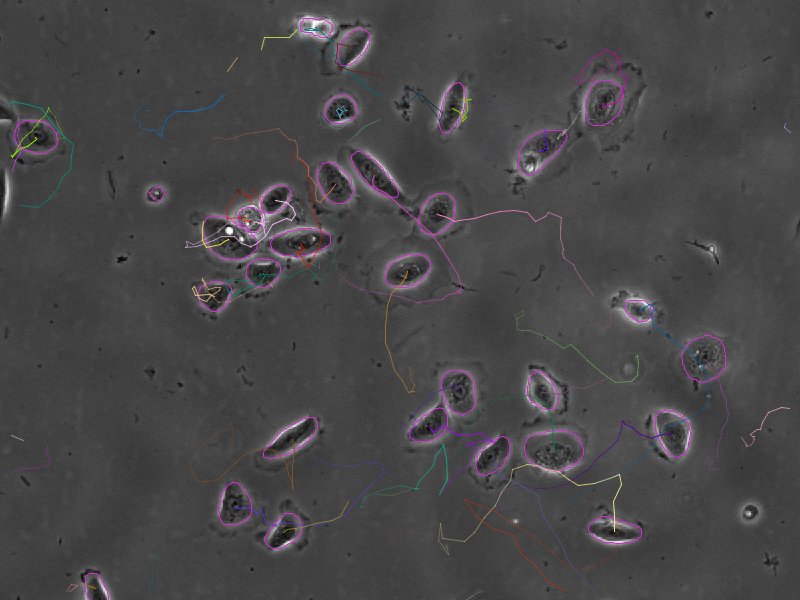}
    \caption{$t=0$}
    \label{fig:exp1}
\end{subfigure}
\hfill
\begin{subfigure}{0.15\textwidth}
    \includegraphics[width=\textwidth]{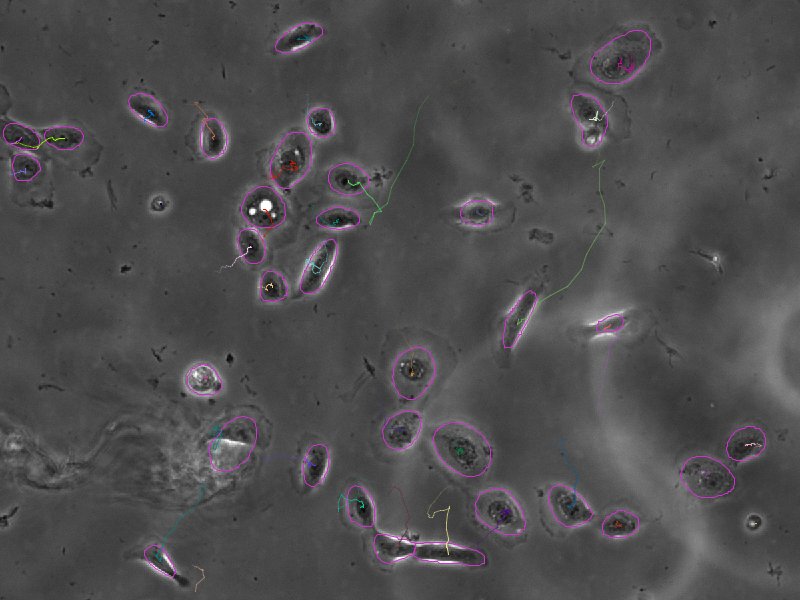}
    \caption{$t=125$ min}
    \label{fig:exp2}
\end{subfigure}
\hfill
\begin{subfigure}{0.15\textwidth}
    \includegraphics[width=\textwidth]{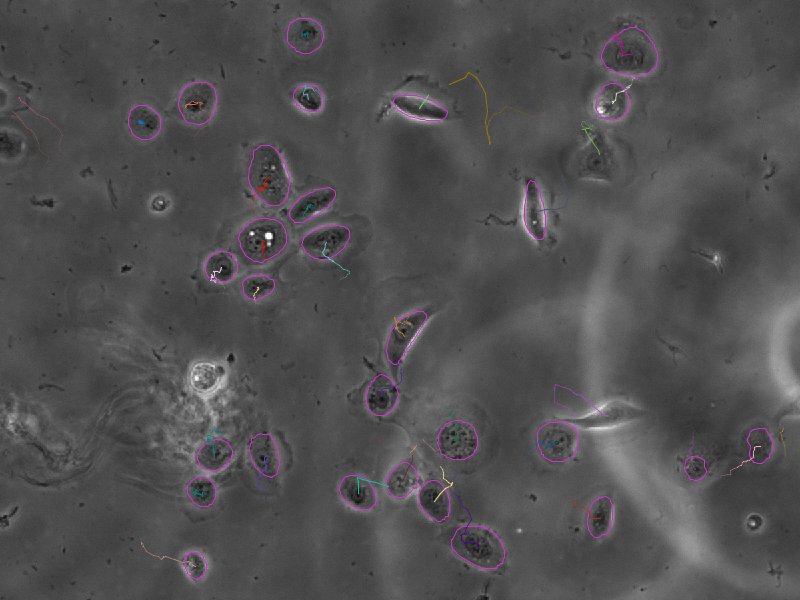}
    \caption{$t=250$ min}
    \label{fig:exp3}
\end{subfigure}
   \hfill
\begin{subfigure}{0.15\textwidth}
    \includegraphics[width=\textwidth]{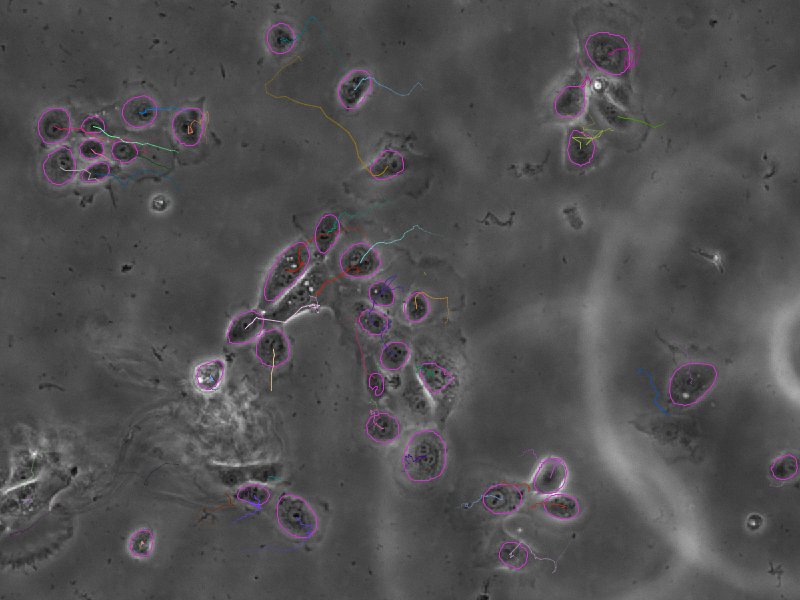}
    \caption{$t=500$ min}
    \label{fig:exp4}
\end{subfigure}
  \hfill
\begin{subfigure}{0.15\textwidth}
    \includegraphics[width=\textwidth]{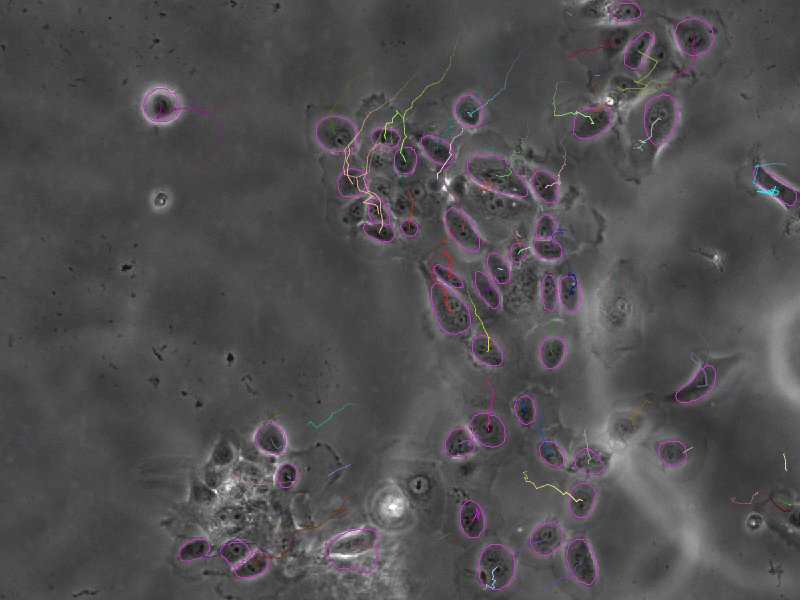}
    \caption{$t=1000 $ min}
    \label{fig:exp5}
\end{subfigure}
  \hfill
\begin{subfigure}{0.15\textwidth}
    \includegraphics[width=\textwidth]{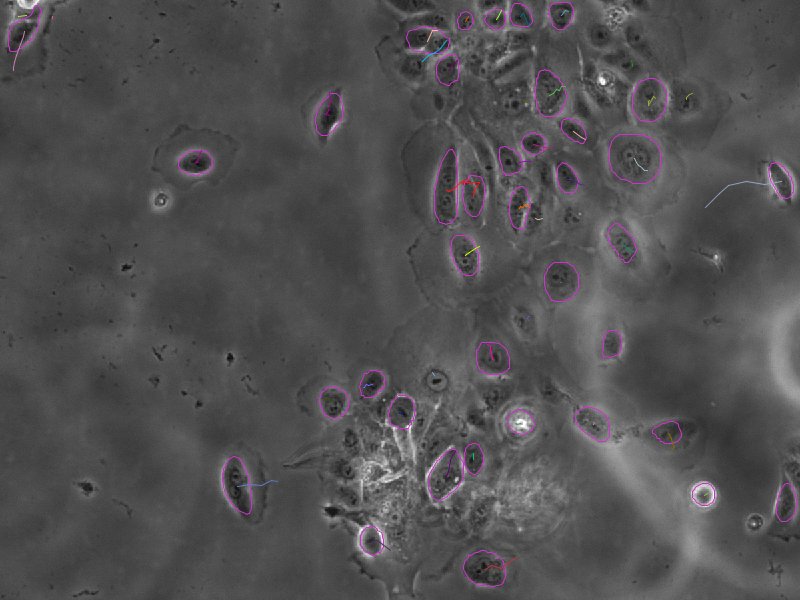}
    \caption{$t=1200$ min}
    \label{fig:exp6}
\end{subfigure}
     
\caption{ One of three samples of experimental observation of Nonmalignant cervical epithelial cell strains which involve change of cell environment due to addition of \texttt{Ca2+} at $t=125$ min. Here the images were acquired every 5 minutes using time lapse microscopy and cell tracking done using \texttt{LAP} tracker in \texttt{Trackmate} plugin of \texttt{FIJI} software
}
\label{fig:experimental}
\end{figure}

The main motivation for this study is that  we want to distinguish whether cells move coherently (form a consistent pattern) or randomly. As an application, the developed technique was used to describe the transition in cellular movement from a disordered state to collective motion when the environment was altered.
We have applied the non-parametric functional hypothesis test based on 1-dimensional topological summary as described in \Cref{sec:hypothesis-pairwise-in-time} on the experimental data of non-malignant cervical epithelial cell strain for detecting the cell aggregation due to an induced change in cell environment as shown in \Cref{fig:experimental}. The time-lapse 2D cell migration dataset obtained by live imaging of lysosensor green stained cells  using Nikon Biostation for every 5 minutes and subjected  them to feature extraction tracker using \texttt{FIJI} software. The migratory status of cells were defined using the \texttt{Trackmate} plugin. From this observed time-lapse  migration data, we used the topological descriptors to capture the time window at which distinct regime changes happens during the migration regardless of variation in the number of cells present in the petri dish for every time frame. \\

\begin{figure}
\centering
\begin{subfigure}{0.15\textwidth}
    \includegraphics[width=\textwidth]{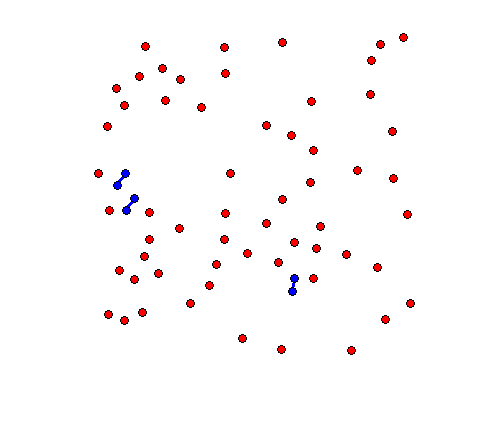}
    \caption{$t=0$}
    \label{fig:exp0}
\end{subfigure}
\hfill
\begin{subfigure}{0.15\textwidth}
    \includegraphics[width=\textwidth]{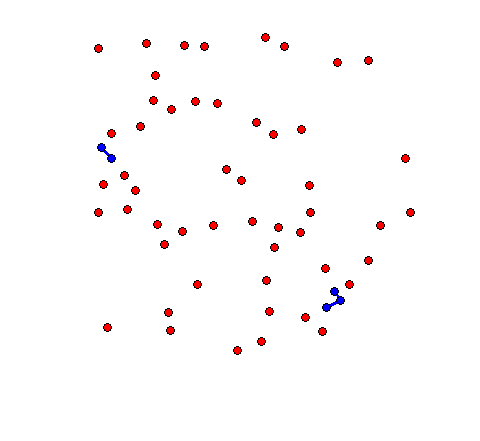}
    \caption{$t=125$ min}
    \label{fig:exp125}
\end{subfigure}
\hfill
\begin{subfigure}{0.15\textwidth}
    \includegraphics[width=\textwidth]{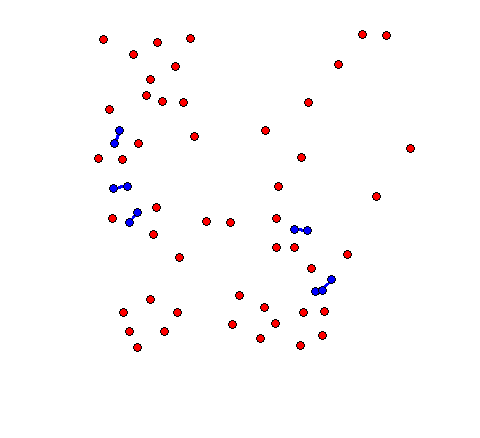}
    \caption{$t=250$ min}
    \label{fig:exp250}
\end{subfigure}
   \hfill
\begin{subfigure}{0.15\textwidth}
    \includegraphics[width=\textwidth]{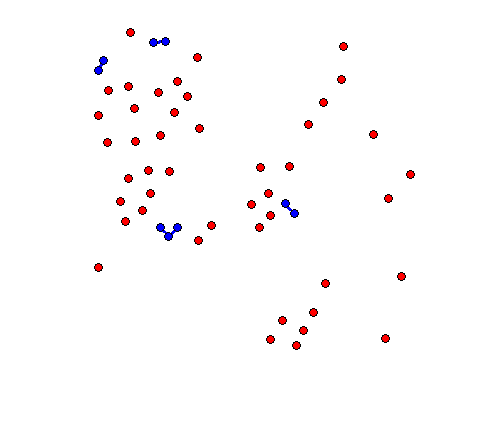}
    \caption{$t=500$ min}
    \label{fig:exp500}
\end{subfigure}
  \hfill
\begin{subfigure}{0.15\textwidth}
    \includegraphics[width=\textwidth]{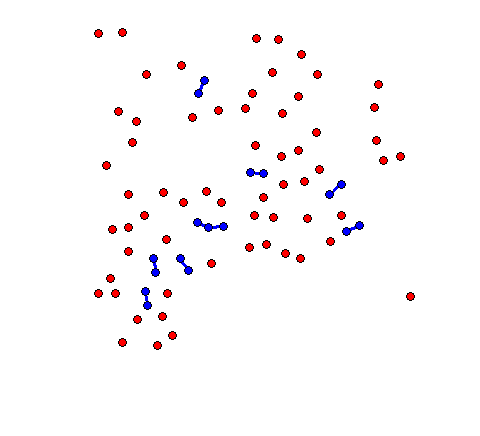}
    \caption{$t=1000 $ min}
    \label{fig:exp1000}
\end{subfigure}
  \hfill
\begin{subfigure}{0.15\textwidth}
    \includegraphics[width=\textwidth]{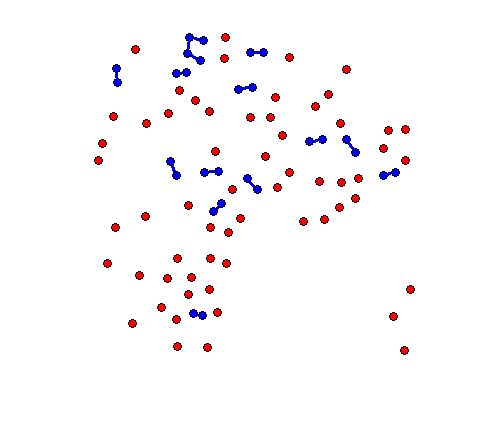}
    \caption{$t=1200$ min}
    \label{fig:exp1200}
\end{subfigure}
     
\caption{ Point cloud data based on the cell positions of considered sample of Nonmalignant cervical epithelial cell strains which involve change of cell environment as in \Cref{fig:experimental}. Here the cell positions are normalized into a domain of dimension $[8 \times 8]$ such that average particles per unit square which is approximately 0.5
}
\label{fig:experimental_point_clouds}
\end{figure}

\begin{figure}
\centering
\begin{subfigure}{0.15\textwidth}
    \includegraphics[width=\textwidth]{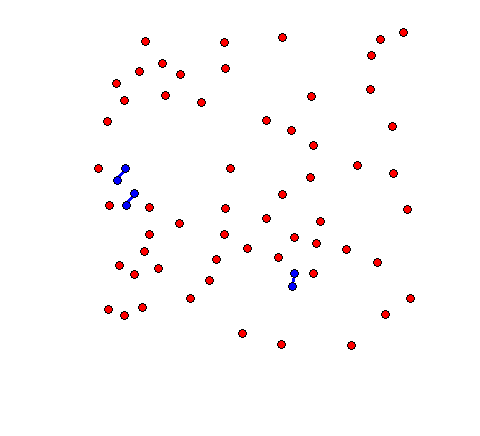}
    \caption{$t=0$}
    \label{fig:Sim_exp_0}
\end{subfigure}
\hfill
\begin{subfigure}{0.15\textwidth}
    \includegraphics[width=\textwidth]{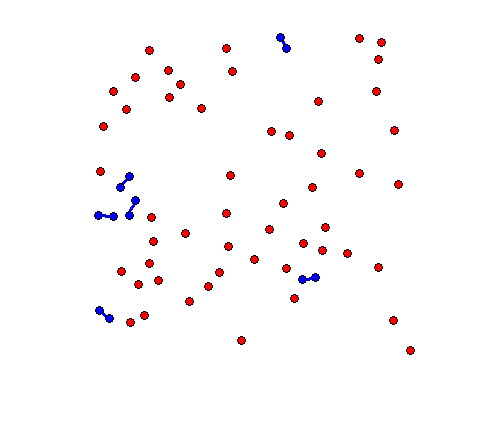}
    \caption{$t=125$ min}
    \label{fig:Sim_exp125}
\end{subfigure}
\hfill
\begin{subfigure}{0.15\textwidth}
    \includegraphics[width=\textwidth]{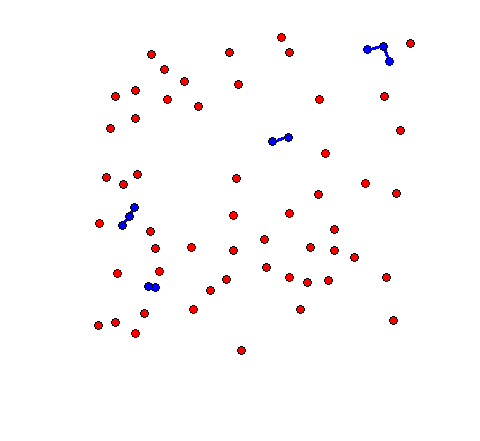}
    \caption{$t=250$ min}
    \label{fig:Sim_exp250}
\end{subfigure}
   \hfill
\begin{subfigure}{0.15\textwidth}
    \includegraphics[width=\textwidth]{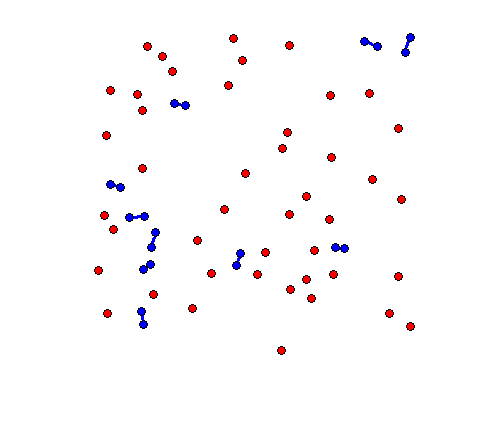}
    \caption{$t=500$ min}
    \label{fig:Sim_exp500}
\end{subfigure}
  \hfill
\begin{subfigure}{0.15\textwidth}
    \includegraphics[width=\textwidth]{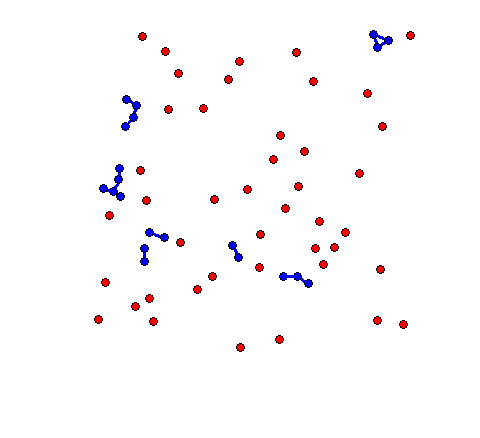}
    \caption{$t=1000 $ min}
    \label{fig:Sim_exp1000}
\end{subfigure}
  \hfill
\begin{subfigure}{0.15\textwidth}
    \includegraphics[width=\textwidth]{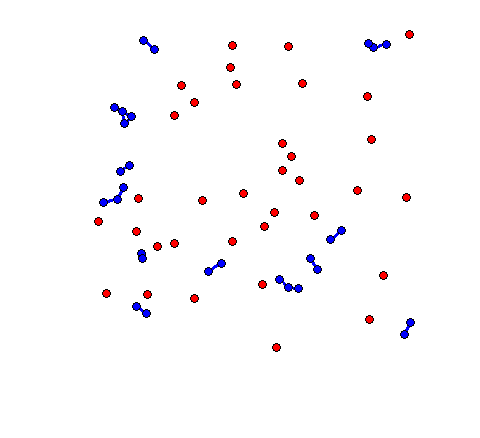}
    \caption{$t=1200$ min}
    \label{fig:Sim_exp1200}
\end{subfigure}
     
\caption{ Point cloud data simulated on the domain of dimension $[8 \times 8]$ with strength of propulsion $\beta = 2.4 \time 10^{-2}$ and strength of adhesion $\alpha = 9 \time 10^{-2}$ which are parameters  corresponding to complete random movement based with initial cell positions as exactly same as the Nonmalignant cervical epithelial cell position at $t=0$ \Cref{fig:exp1} and the evolve over the time period of $t=1225$ min
}
\label{fig:Sim_exp_point_clouds}
\end{figure}

Using the test described in \Cref{sec:pointwise-two-time-test}, we compared between the point cloud data generated from this observed time-lapse  migration given in \Cref{fig:Sim_exp_point_clouds} and the point cloud data generated based on the parameters ($\alpha= 9 \times 10^{-2}, \beta= 2.4 \times 10^{-2}$) corresponding to complete random movement from collection motion model \Cref{eq:dorsogna-model} at every time frame. Note that the simulations in \Cref{fig:Sim_exp_point_clouds} consist of initial positions as similar to the point cloud data of \Cref{fig:experimental} at $t=0$. Further each sample of the point cloud data corresponding to experiment (\Cref{fig:experimental_point_clouds}) and the parallel random movement simulation (\Cref{fig:Sim_exp_point_clouds}) is set to have same average particles per unit square which is approximately 0.5. Through this way we can identify whether the cell migration is completely random or whether and when the change of cell environment due to addition of \texttt{Ca2+} has an effect on driving the randomly moving cells towards aggregation phase.

\begin{figure}
\centering
\begin{subfigure}{0.15\textwidth}
    \includegraphics[width=\textwidth]{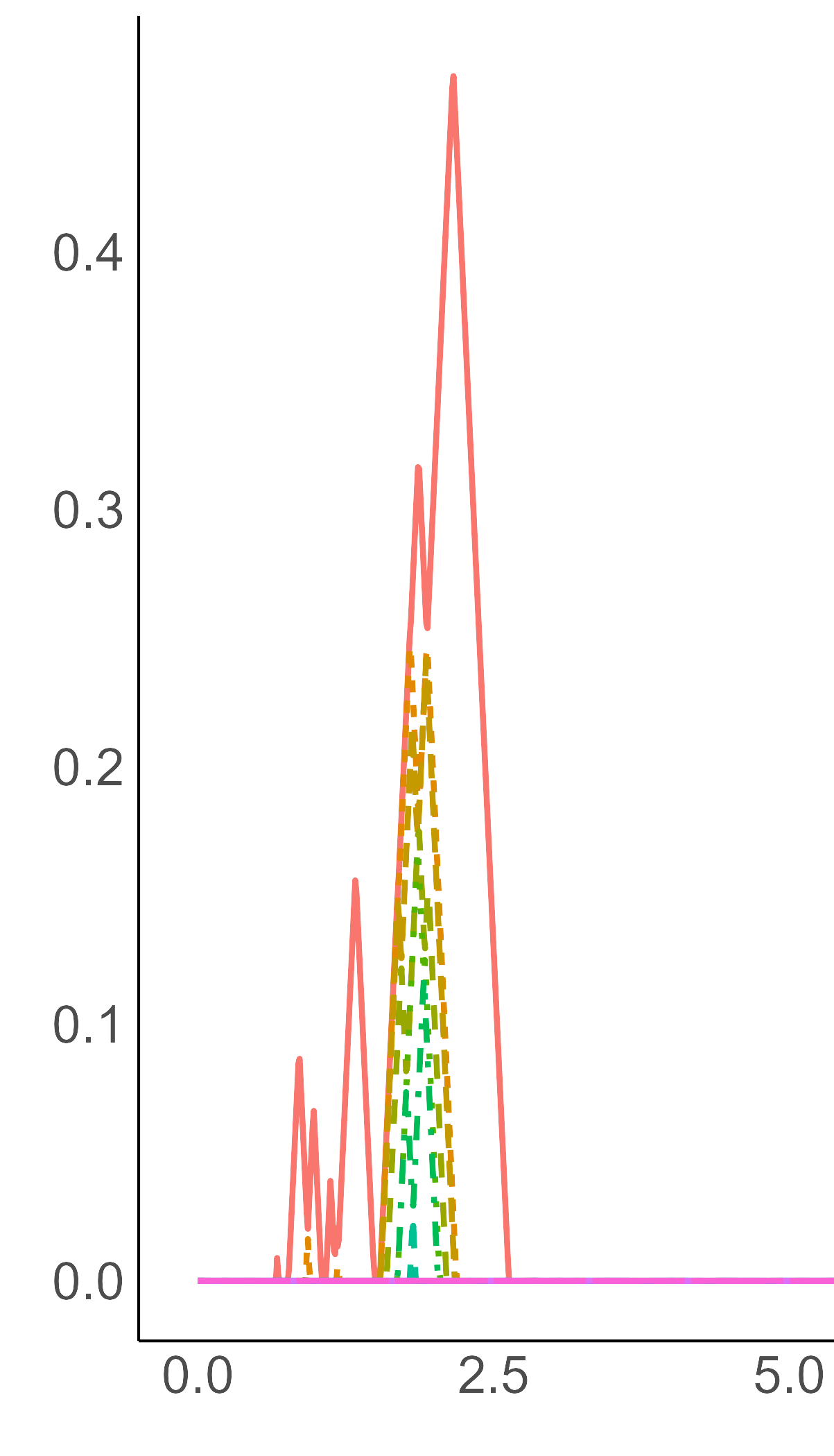}
    \caption{$t=0$}
    \label{fig:Exp_time_frame0}
\end{subfigure}
\hfill
\begin{subfigure}{0.15\textwidth}
    \includegraphics[width=\textwidth]{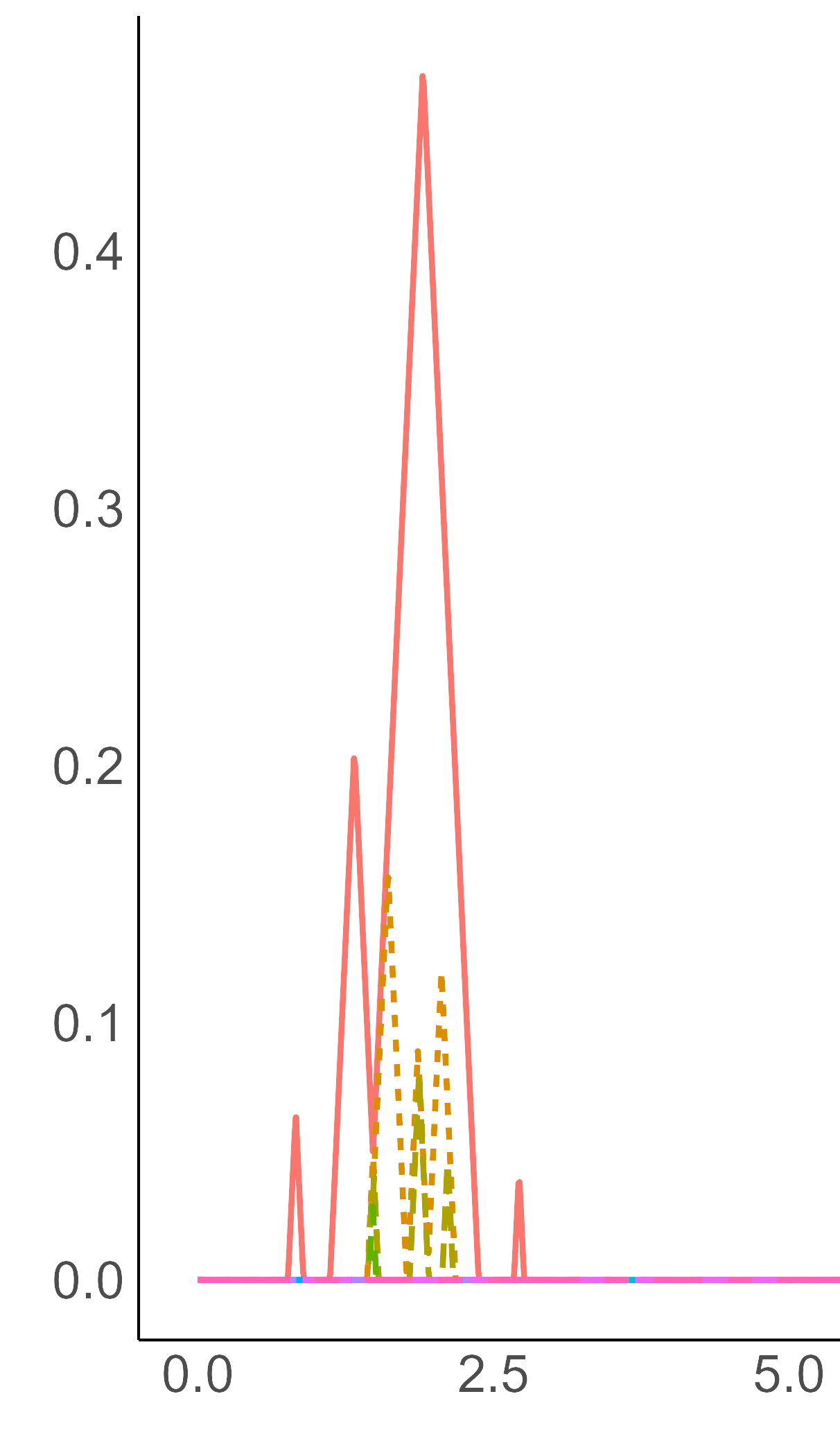}
    \caption{$t=125$ min}
    \label{fig:Exp_time_frame125}
\end{subfigure}
\hfill
\begin{subfigure}{0.15\textwidth}
    \includegraphics[width=\textwidth]{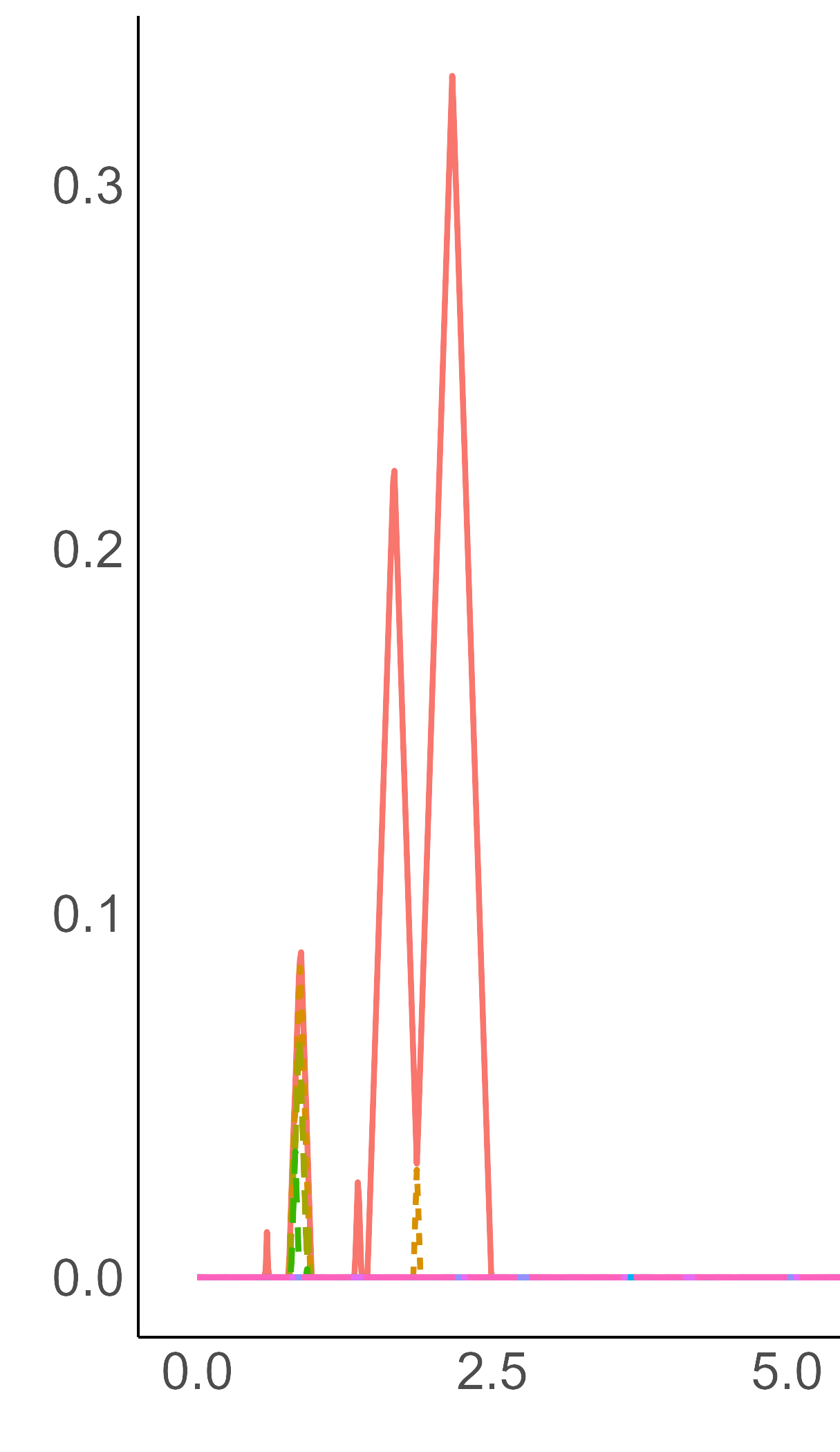}
    \caption{$t=250$ min}
    \label{fig:Exp_time_frame250}
\end{subfigure}
   \hfill
\begin{subfigure}{0.15\textwidth}
    \includegraphics[width=\textwidth]{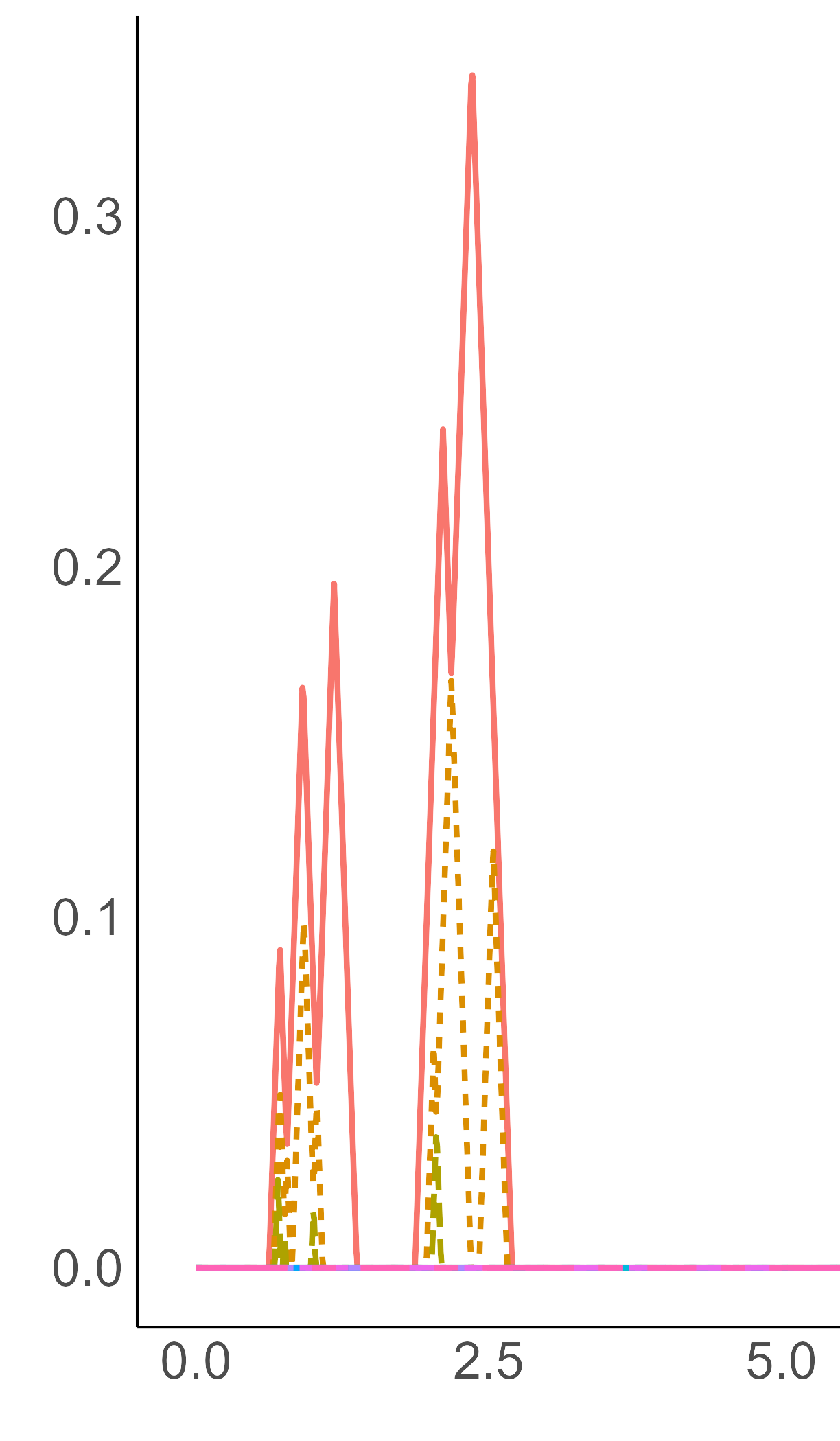}
    \caption{$t=500$ min}
    \label{fig:Exp_time_frame500}
\end{subfigure}
  \hfill
\begin{subfigure}{0.15\textwidth}
    \includegraphics[width=\textwidth]{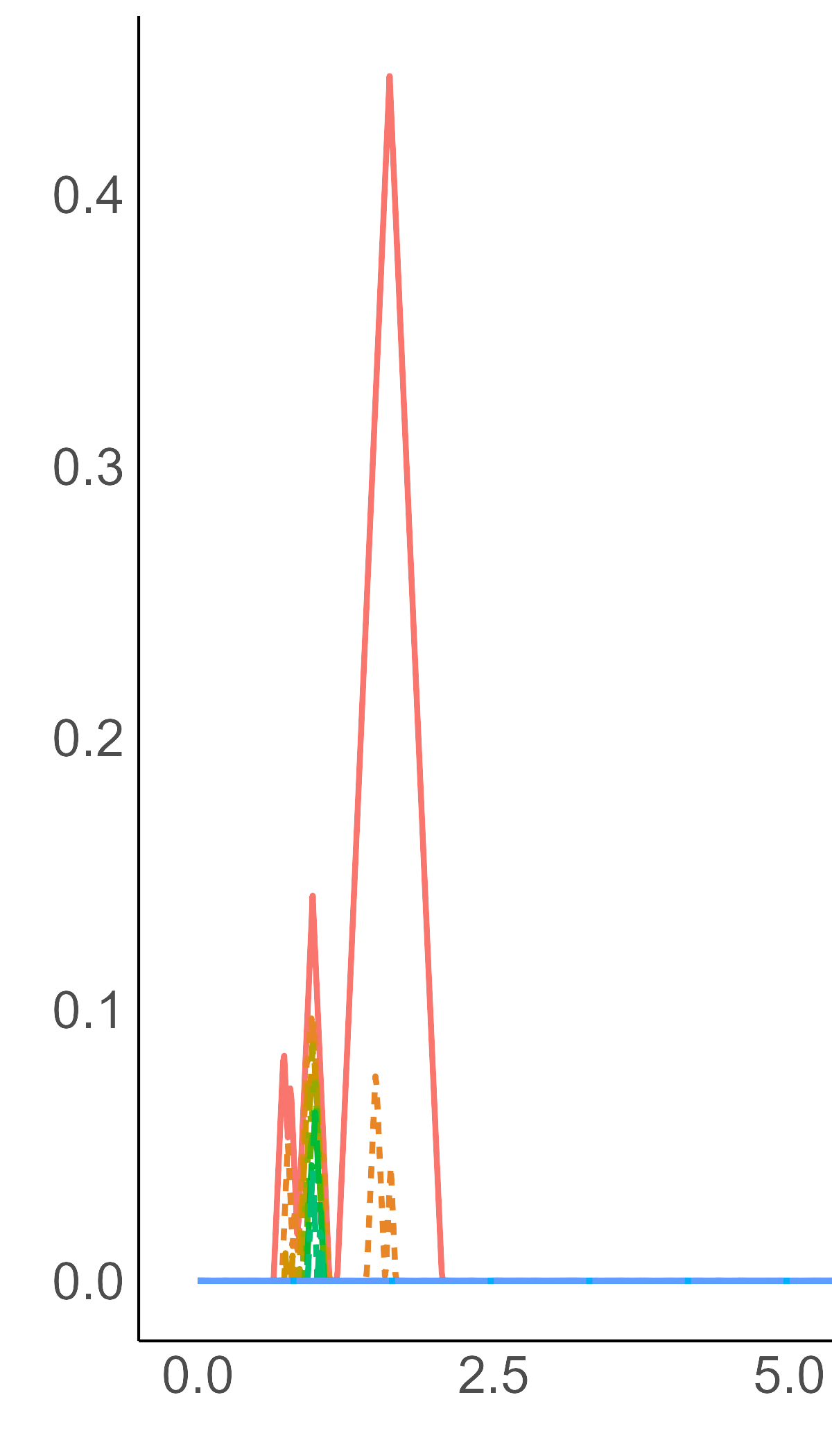}
    \caption{$t=1000 $ min}
    \label{fig:Exp_time_frame1000}
\end{subfigure}
  \hfill
\begin{subfigure}{0.15\textwidth}
    \includegraphics[width=\textwidth]{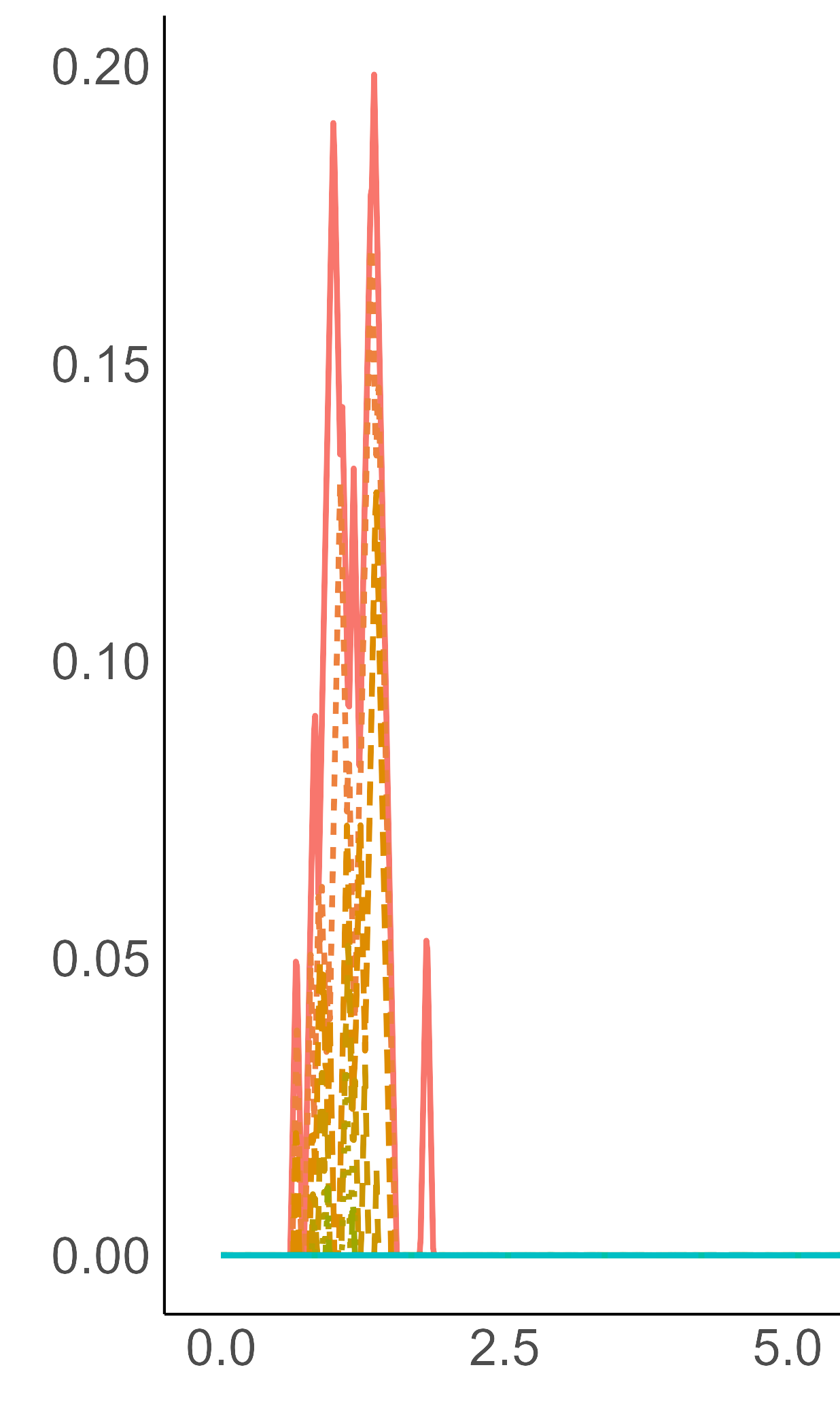}
    \caption{$t=1200$ min}
    \label{fig:Exp_time_frame1200}
\end{subfigure}
     
\caption{ One dimensional Persistence landscape corresponding to the point cloud data  \Cref{fig:experimental_point_clouds}
}
\label{fig:experimental_PLs1}
\end{figure}

\begin{figure}
\centering
\begin{subfigure}{0.15\textwidth}
    \includegraphics[width=\textwidth]{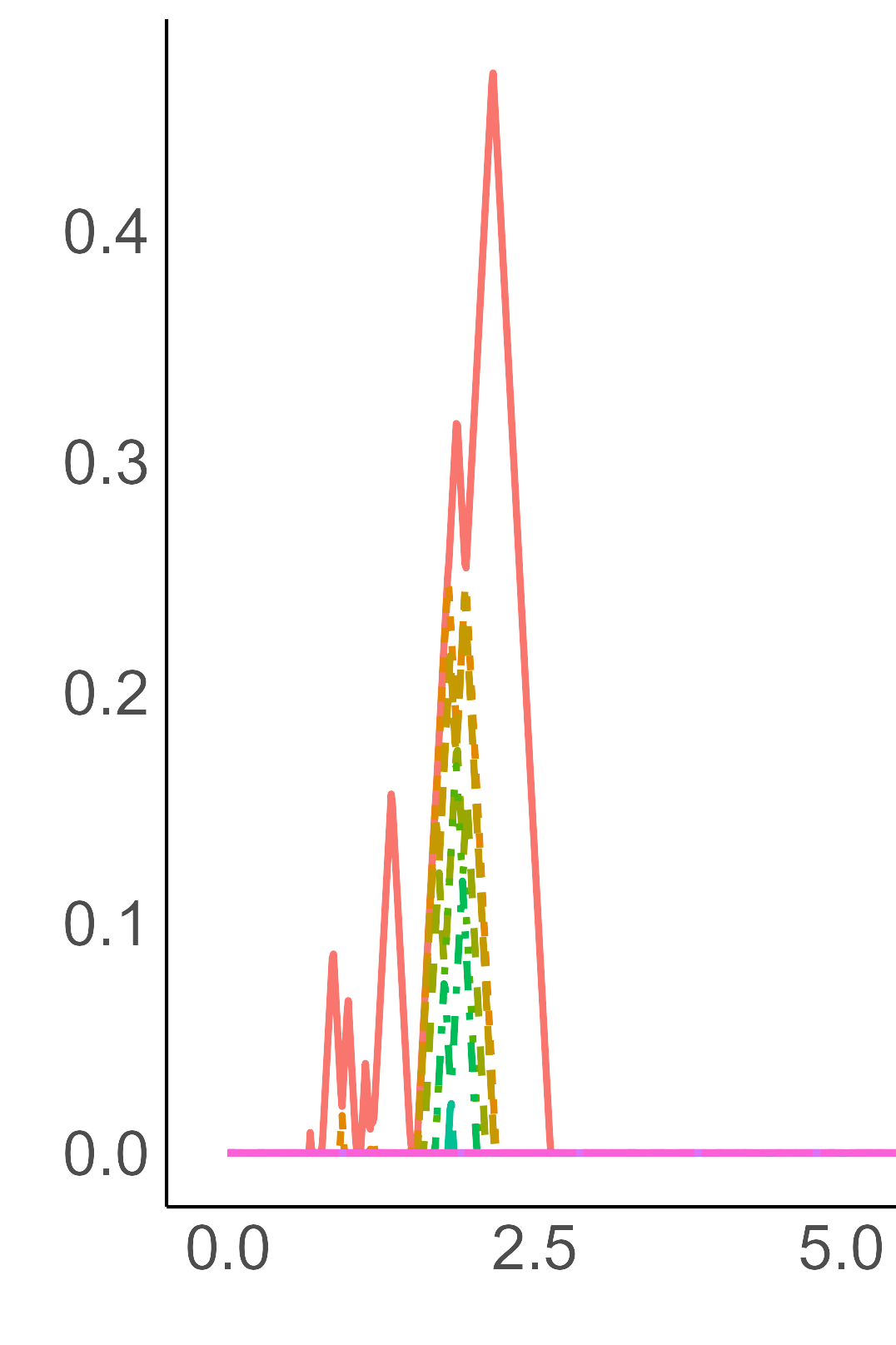}
    \caption{$t=0$}
    \label{fig:Sim_exp_time_frame_0}
\end{subfigure}
\hfill
\begin{subfigure}{0.15\textwidth}
    \includegraphics[width=\textwidth]{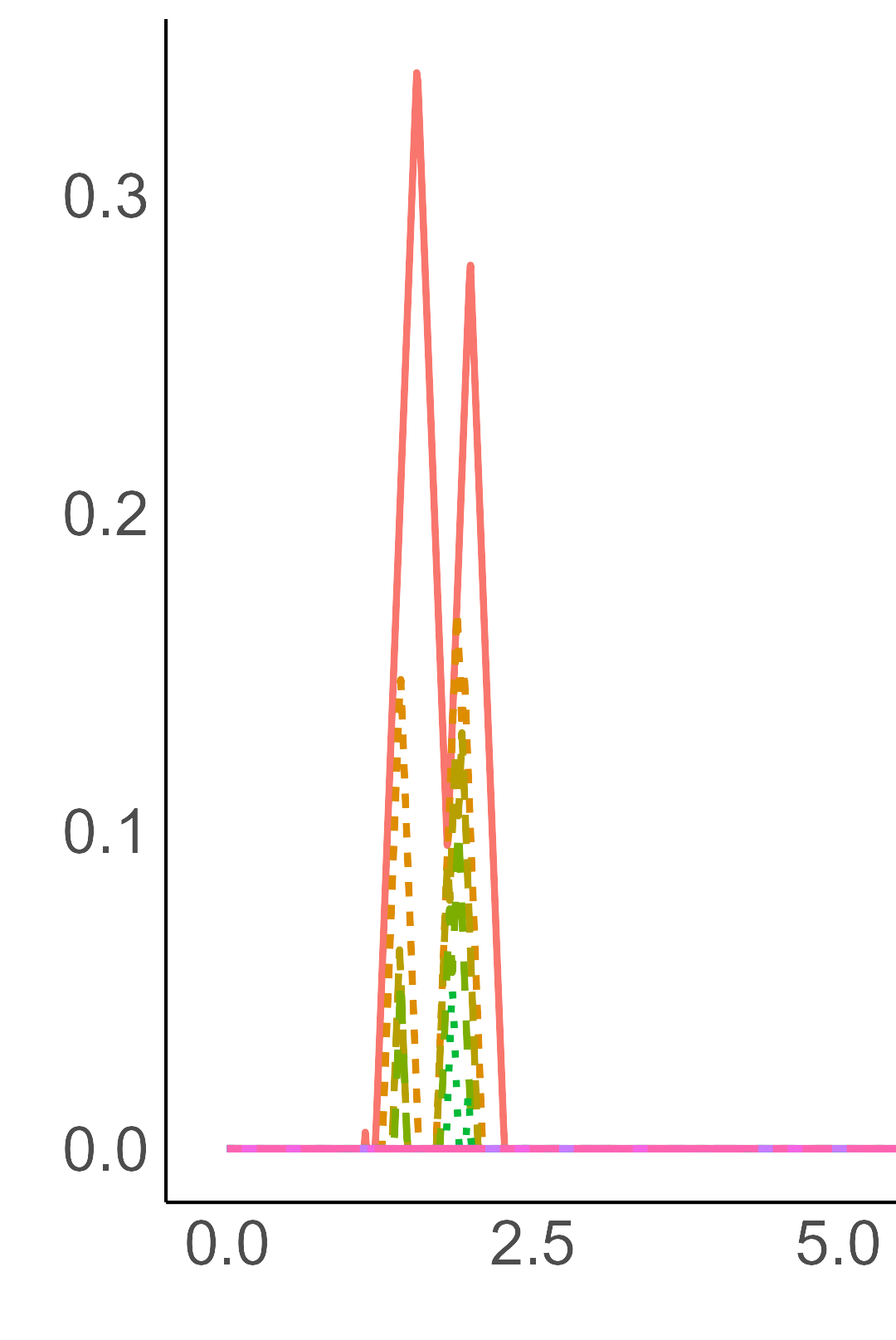}
    \caption{$t=125$ min}
    \label{fig:Sim_exp_time_frame25}
\end{subfigure}
\hfill
\begin{subfigure}{0.15\textwidth}
    \includegraphics[width=\textwidth]{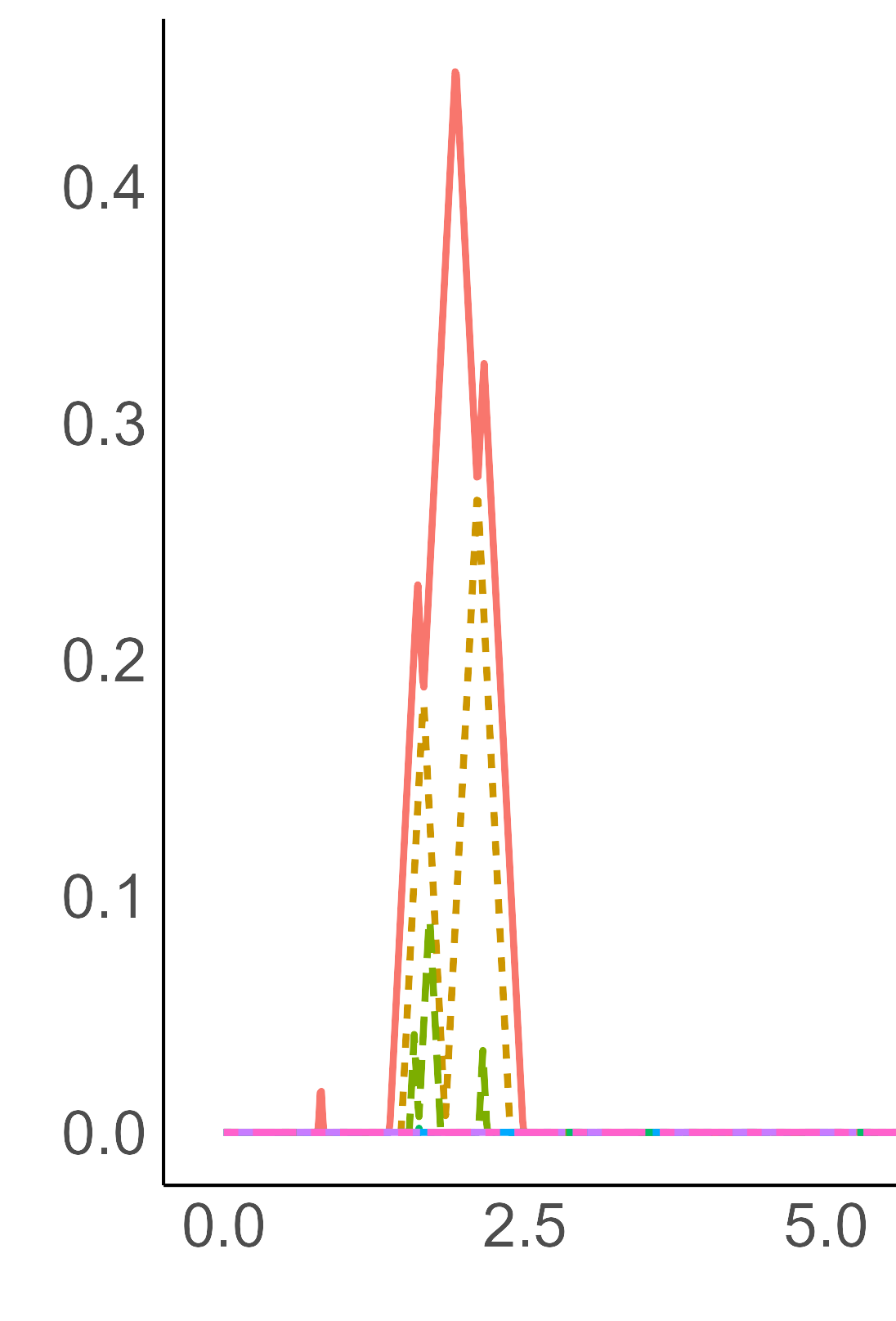}
    \caption{$t=250$ min}
    \label{fig:Sim_exp_time_frame50}
\end{subfigure}
   \hfill
\begin{subfigure}{0.15\textwidth}
    \includegraphics[width=\textwidth]{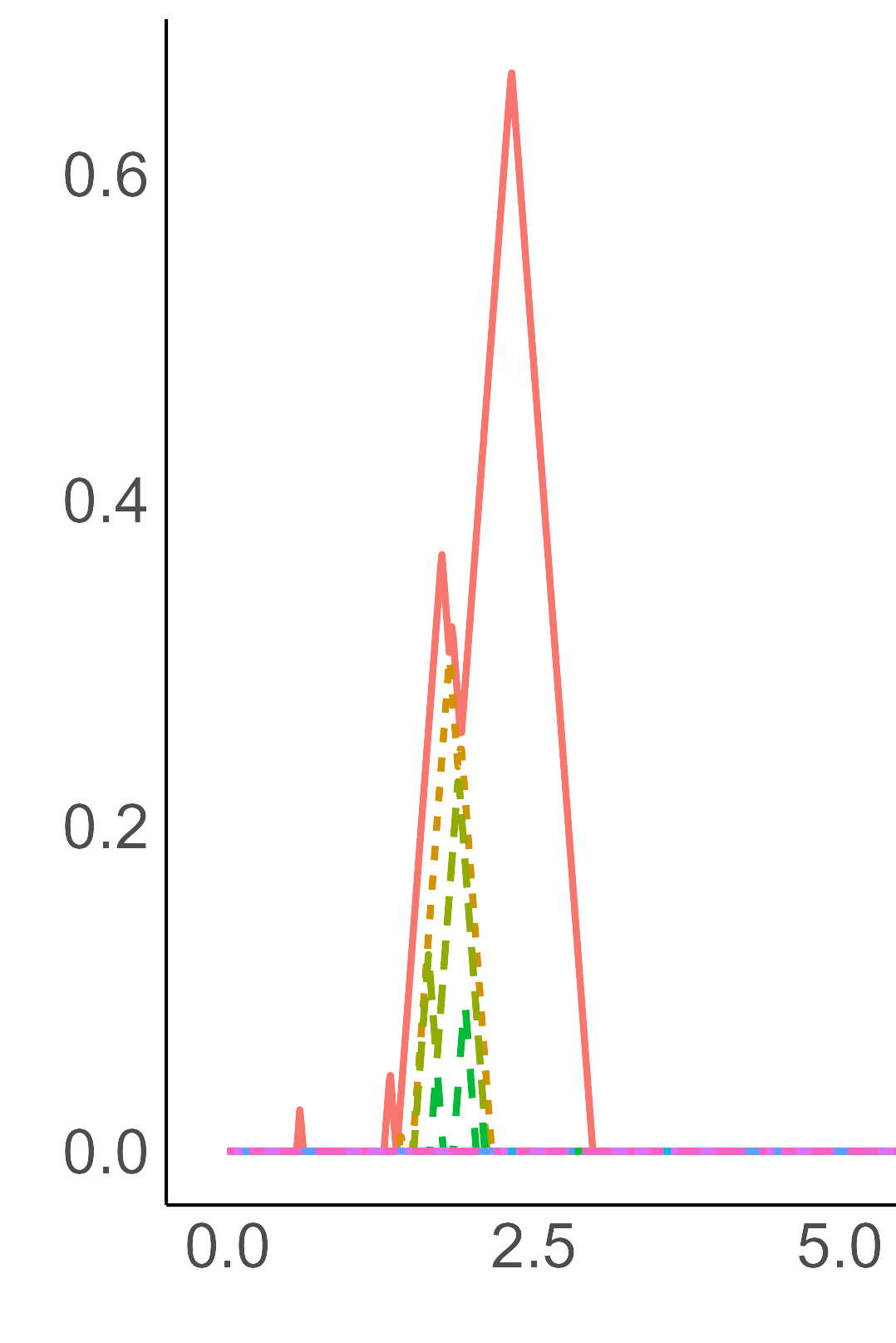}
    \caption{$t=500$ min}
    \label{fig:Sim_exp_time_frame100}
\end{subfigure}
  \hfill
\begin{subfigure}{0.15\textwidth}
    \includegraphics[width=\textwidth]{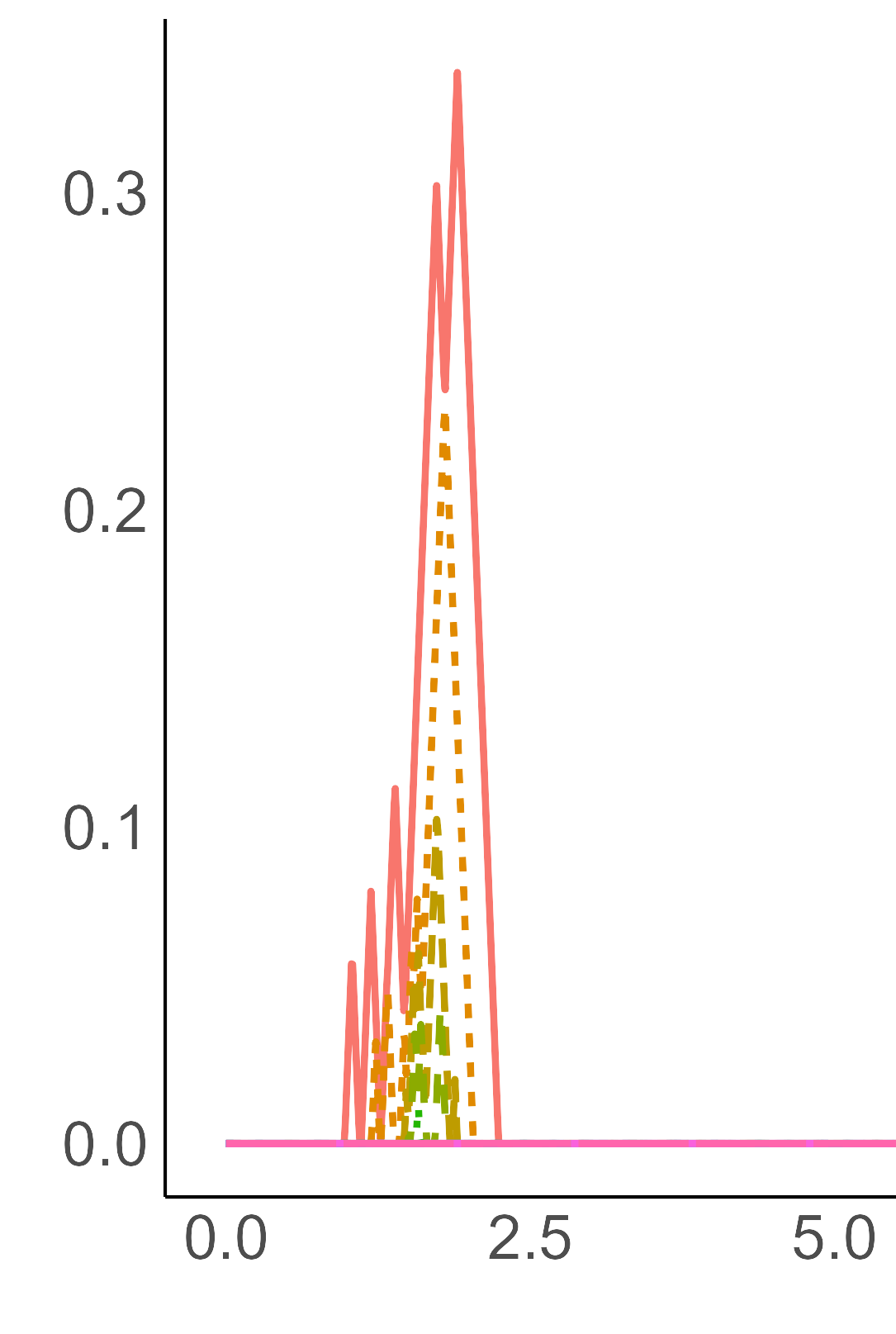}
    \caption{$t=1000 $ min}
    \label{fig:Sim_exp_time_frame200}
\end{subfigure}
  \hfill
\begin{subfigure}{0.15\textwidth}
    \includegraphics[width=\textwidth]{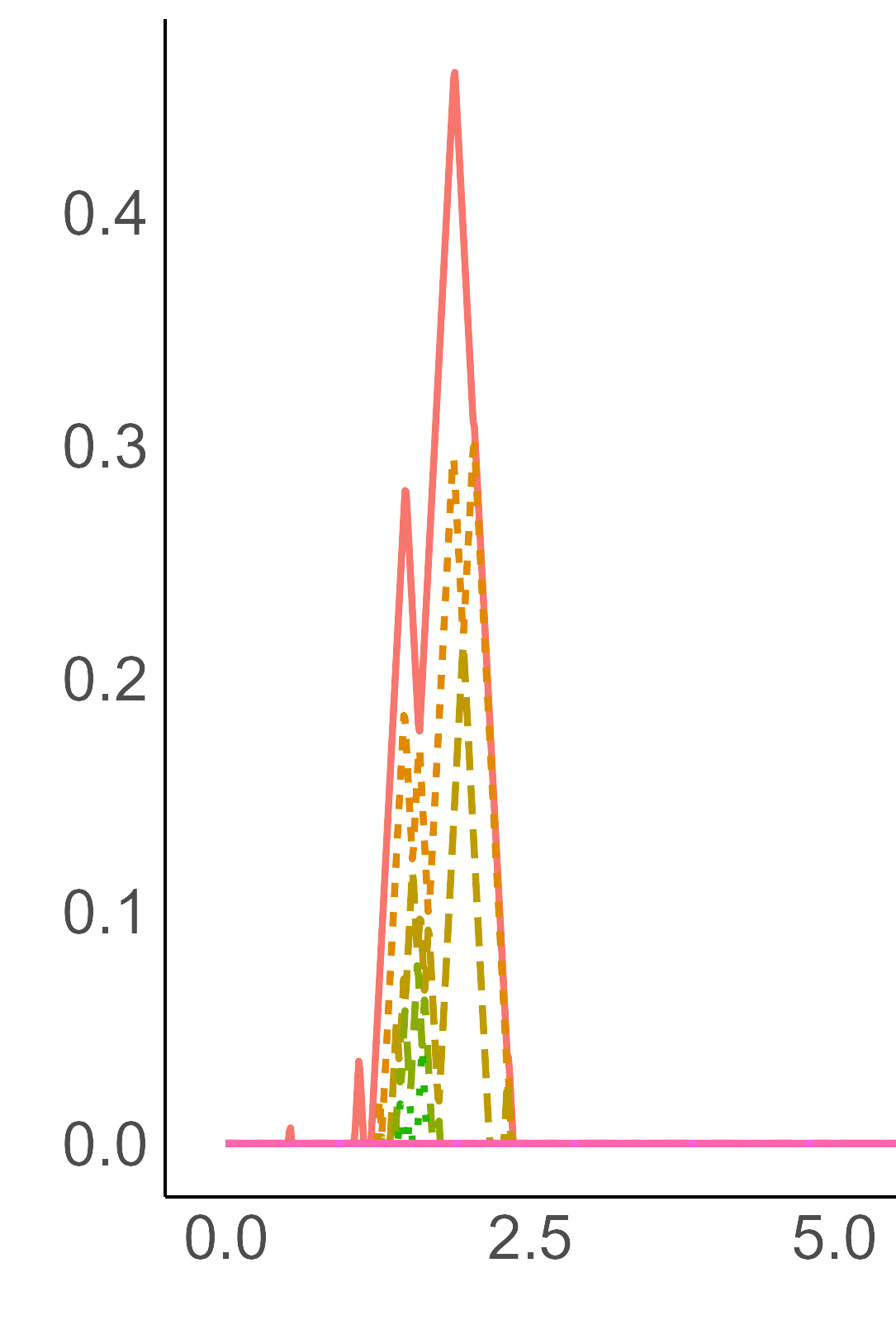}
    \caption{$t=1200$ min}
    \label{fig:Sim_exp_time_frame240}
\end{subfigure}
     
\caption{ One dimensional Persistence landscape corresponding to the point cloud data  \Cref{fig:Sim_exp_point_clouds}
}
\label{fig:Sim_exp_PLs}
\end{figure}

\begin{figure}
    \includegraphics[width=12cm]{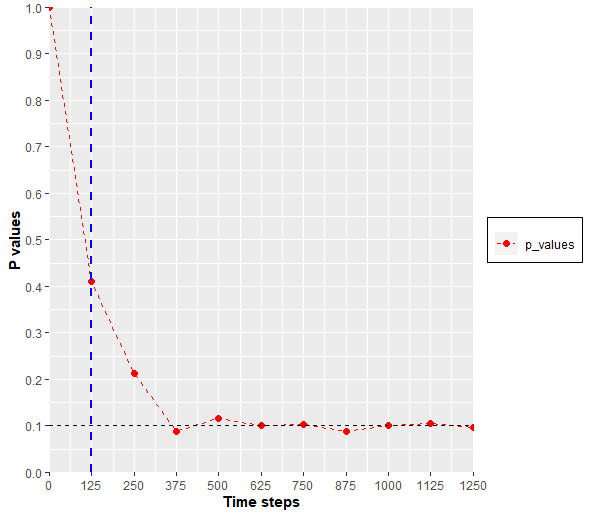}
    \caption{The minimum adjusted p-values at each time step obtained from the tests which are implemented to compare the of simulations between two groups of namely the point clouds corresponding to experimental observations and (\Cref{fig:experimental_point_clouds}) and point clouds corresponding to random movement (\Cref{fig:Sim_exp_point_clouds}). The blue line indicates the exact time $t=125$ at which \texttt{Ca2+} is added. At the level of significance \(\alpha=0.10\) indicated by horizontal black dotted line, we can \emph{reject} the null hypothesis \(H_{0}\)}
    \label{fig:Experimental_result}
\end{figure}

The main result of comparing between experimental vs. simulation \Cref{fig:Experimental_result} shows that there is a decline in \emph{p}-values over the time interval and thereby detects significant deviation of in the topological distribution of experimental cell point cloud data from the generated random point cloud data after the time point $t=130$ minute. We can note that the significant effect of the addition of \texttt{Ca2+} on driving the randomly moving cells towards aggregation phase can be clearly quantified though the statistically significant differences between the topological distribution of PLC descriptors.

\FloatBarrier
\section{Limitations}
\label{limitation}
When the underlying change is topological in nature, the suggested statistical inference approach can detect if the system of particles regime change based on suite of simulation experiments and an experimental application. Despite these results, there are certain limitations that can be improved as follows:

\begin{itemize}

\item \emph{Different dynamical systems displaying collective behaviors:} We have employed the first order discrete stochastic model describing the motion of "cells" (particles) \cites{bhaskarTopologicalDataAnalysis2021} to study whether our approach enables us to detect when the regime shifts in an (experimental) real world system occurs. It was demonstrated that the proposed statistical inference method with PLCs is effective for identifying the different dynamical transitions that the system undergoes. We can extend to detect the time of transition of coherent movements towards different pattern configurations such as flock, mills, rings exhibited through second-order swarming systems involving self-propulsion and potential terms.

\item \emph{Computation of homology:}  We have used the common method for constructing a simplicial complex to calculate persistent homology, Vietoris-Rips complex. However there are different methods to construct a simplicial complex on a given point cloud  such as alpha complexes. It is predicted in ~\cite{Benchmarking} that based on run time complexity alpha complexes are better for low dimensional point clouds and Vietoris-Rips are better for fewer points in higher dimension. Therefore, depending on the dimension of the point cloud we can use the appropriate simplicial complex. Another popular approach in TDA is to convert a point-cloud of data into a function using Kernel density estimate ~\cite{Fasy} or Distance to measure functions ~\cite{Chazal} which is used to compute the persistent homology through finding superlevel sets. This method is less sensitive to noisy observations.

\end{itemize}

\FloatBarrier
\section{Conclusion and future works}
\label{conclusion_future}
This study presents a  novel approach of using the non-parametric functional significance test with PLCs to detect the topological difference between snapshots of particle configuration for the purpose of identifying  when the transition in pattern regimes likely to happen during an observed period of migration of system of particles. This method identifies the impact of distribution differences in topological features in detecting the pattern regime change. These specific results about the collective motion model results presented demonstrate that this method can also be used to perform the task of retrieval of the time instances when the aggregating behaviour seen in terms of topological features for the context of wound healing or cancer cell tumor migration. We also want to emphasize that the apparent significant reduction in information, occurring during the transition from a complex cell migration to its persistent homology, still preserves a remarkable amount of information about the underlying dynamical process. Given its inherently discrete nature, the topological approach holds significant potential for analyzing and establishing connections between experimental data and numerical models. Furthermore, the following are the ways to improve our method with to adapt for experimental data
\begin{itemize}
    \item Use of appropriate sampling technique to generate adequate samples of cell configuration from a single experimental data
    \item Overcome the variation in capturing transition time frame due to changes in number of particles with the use of kernel density estimate

\end{itemize}

\FloatBarrier

\bibliography{references}

\begin{bibdiv}
\begin{biblist}

\bib{SelforganizationFundamentCell}{misc}{
       title={Self-organization: The fundament of cell biology |
  {{Philosophical Transactions}} of the {{Royal Society B}}: {{Biological
  Sciences}}},
         how={https://royalsocietypublishing.org/doi/10.1098/rstb.2017.0103},
}

\bib{adamsTopologyAppliedMachine2021}{article}{
      author={Adams, Henry},
      author={Moy, Michael},
       title={Topology {{Applied}} to {{Machine Learning}}: {{From Global}} to
  {{Local}}},
    language={English},
        date={2021},
        ISSN={2624-8212},
     journal={Frontiers in Artificial Intelligence},
      volume={0},
}

\bib{bhaskarTopologicalDataAnalysis2021}{article}{
      author={Bhaskar, Dhananjay},
      author={Zhang, William~Y.},
      author={Wong, Ian~Y.},
       title={Topological data analysis of collective and individual epithelial
  cells using persistent homology of loops},
        date={2021-05},
        ISSN={1744-6848},
     journal={Soft Matter},
      volume={17},
      number={17},
       pages={4653\ndash 4664},
}

\bib{Wilcoxon}{article}{
      author={Blair, R.~Clifford},
      author={Higgins, James~J.},
       title={A comparison of the power of wilcoxon's rank-sum statistic to
  that of student's t statistic under various nonnormal distributions},
        date={1980},
        ISSN={03629791},
     journal={Journal of Educational Statistics},
      volume={5},
      number={4},
       pages={309\ndash 335},
         url={http://www.jstor.org/stable/1164905},
}

\bib{bubenikStatisticalTopologicalData}{article}{
      author={Bubenik, Peter},
       title={Statistical {{Topological Data Analysis}} using {{Persistence
  Landscapes}}},
    language={en},
       pages={26},
}

\bib{bubenikPersistenceLandscapesToolbox2017}{article}{
      author={Bubenik, Peter},
      author={D{\l}otko, Pawe{\l}},
       title={A persistence landscapes toolbox for topological statistics},
        date={2017-01},
        ISSN={07477171},
     journal={Journal of Symbolic Computation},
      volume={78},
       pages={91\ndash 114},
}

\bib{bukkuriApplicationsTopologicalData2021}{article}{
      author={Bukkuri, Anuraag},
      author={Andor, Noemi},
      author={Darcy, Isabel~K.},
       title={Applications of {{Topological Data Analysis}} in {{Oncology}}},
        date={2021-04},
        ISSN={2624-8212},
     journal={Frontiers in Artificial Intelligence},
      volume={4},
       pages={659037},
}

\bib{Chazal}{article}{
      author={Chazal, Fr{\'e}d{\'e}ric},
      author={Fasy, Brittany},
      author={Lecci, Fabrizio},
      author={Bertr},
      author={Michel},
      author={Aless},
      author={ro~Rinaldo},
      author={Wasserman, Larry},
       title={Robust topological inference: Distance to a measure and kernel
  distance},
        date={2018},
     journal={Journal of Machine Learning Research},
      volume={18},
      number={159},
       pages={1\ndash 40},
         url={http://jmlr.org/papers/v18/15-484.html},
}

\bib{Westfall}{article}{
      author={Cox, Dennis~D.},
      author={Lee, Jong~Soo},
       title={Pointwise testing with functional data using the westfall young
  randomization method},
        date={2008},
        ISSN={00063444, 14643510},
     journal={Biometrika},
      volume={95},
      number={3},
       pages={621\ndash 634},
         url={http://www.jstor.org/stable/20441490},
}

\bib{dorsognaSelfPropelledParticlesSoftCore2006}{article}{
      author={D'Orsogna, M.~R.},
      author={Chuang, Y.~L.},
      author={Bertozzi, A.~L.},
      author={Chayes, L.~S.},
       title={Self-{{Propelled Particles}} with {{Soft}}-{{Core Interactions}}:
  Patterns, {{Stability}}, and {{Collapse}}},
        date={2006-03},
        ISSN={0031-9007, 1079-7114},
     journal={Physical Review Letters},
      volume={96},
      number={10},
       pages={104302},
}

\bib{edelsbrunnerTopologicalPersistenceSimplification2002}{article}{
      author={{Edelsbrunner}},
      author={{Letscher}},
      author={{Zomorodian}},
       title={Topological {{Persistence}} and {{Simplification}}},
    language={en},
        date={2002-11},
        ISSN={0179-5376, 1432-0444},
     journal={Discrete \& Computational Geometry},
      volume={28},
      number={4},
       pages={511\ndash 533},
}

\bib{edelsbrunner2008}{book}{
      author={Edelsbrunner, Herbert},
      author={Harer, John},
       title={Persistent homology\textemdash a survey},
   publisher={{American Mathematical Society}},
        date={2008},
      volume={453},
}

\bib{Edelsbrunner2010}{book}{
      author={Edelsbrunner, Herbert},
      author={Harer, John},
       title={Computational topology an introduction},
   publisher={American Mathematical Soc.},
        date={2010},
}

\bib{Benchmarking}{article}{
      author={et~al, Somasundaram~EV},
       title={Benchmarking r packages for calculation of persistent homology},
        date={2021},
     journal={R J},
       pages={184\ndash 193},
}

\bib{Fasy}{article}{
      author={Fasy, et~al, Brittany~Terese},
       title={Confidence sets for persistence diagrams},
        date={2014},
     journal={The Annals of Statistics},
      volume={42},
      number={6},
       pages={2301\ndash 2339},
         url={http://www.jstor.org/stable/43556495},
}

\bib{ghristBarcodesPersistentTopology2008}{article}{
      author={Ghrist, Robert},
       title={Barcodes: {{The}} persistent topology of data},
    language={en},
        date={2008},
        ISSN={0273-0979, 1088-9485},
     journal={Bulletin of the American Mathematical Society},
      volume={45},
      number={1},
       pages={61\ndash 75},
}

\bib{NonparametricFDA}{article}{
      author={González-Manteiga, W. F.~Ferraty},
      author={Vieu, P.},
       title={Nonparametric functional data analysis: theory and practice},
        date={2008},
     journal={Computational Statistics},
      volume={23},
       pages={341\ndash 342},
         url={https://doi.org/10.1007/s00180-008-0111-2},
}

\bib{Hollander2013}{book}{
      author={Hollander, Myles},
      author={Wolfe, Douglas~A.},
      author={Chicken, Eric},
       title={Nonparametric statistical methods},
     edition={Third edition},
   publisher={John Wiley \& Sons, Inc.},
     address={Hoboken, New Jersey},
        date={2013},
}

\bib{Ramsay}{book}{
      author={J.O.~Ramsay, Spencer~Graves, Giles~Hooker},
       title={{Functional Data Analysis with R and MATLAB}},
   publisher={Springer},
        date={2009},
}

\bib{Keser2014ComparingTM}{article}{
      author={Keser, Istem~Koymen},
       title={Comparing two mean humidity curves using functiona t-tests:
  Turkey case},
        date={2014},
     journal={Electronic Journal of Applied Statistical Analysis},
      volume={7},
       pages={254\ndash 278},
         url={https://api.semanticscholar.org/CorpusID:55039908},
}

\bib{kovacev-nikolicUsingPersistentHomology2016}{article}{
      author={{Kovacev-Nikolic}, Violeta},
      author={Bubenik, Peter},
      author={Nikoli{\'c}, Dragan},
      author={Heo, Giseon},
       title={Using persistent homology and dynamical distances to analyze
  protein binding},
        date={2016-01},
        ISSN={1544-6115, 2194-6302},
     journal={Statistical Applications in Genetics and Molecular Biology},
      volume={15},
      number={1},
      eprint={1412.1394},
}

\bib{Medina2016StatisticalMI}{article}{
      author={Medina, Patrick~S},
      author={Doerge, Rebecca~W.},
       title={Statistical methods in topological data analysis for complex,
  high-dimensional data},
        date={2016},
     journal={arXiv: Applications},
}

\bib{Sign_Wilcoxon_and_Mann-Whitney}{article}{
      author={Meléndez, Rafael},
      author={Giraldo, Ramón},
      author={Leiva, Víctor},
       title={Sign, wilcoxon and mann-whitney tests for functional data: An
  approach based on random projections},
        date={2021},
        ISSN={2227-7390},
     journal={Mathematics},
      volume={9},
      number={1},
         url={https://www.mdpi.com/2227-7390/9/1/44},
}

\bib{mileykoProbabilityMeasuresSpace2011}{article}{
      author={Mileyko, Yuriy},
      author={Mukherjee, Sayan},
      author={Harer, John},
       title={Probability measures on the space of persistence diagrams},
    language={en},
        date={2011-11},
        ISSN={0266-5611},
     journal={Inverse Problems},
      volume={27},
      number={12},
       pages={124007},
}

\bib{oudotPersistenceTheoryQuiver2015}{book}{
      author={Oudot, Steve},
       title={Persistence {{Theory}}: {{From Quiver Representations}} to {{Data
  Analysis}}},
      series={Mathematical {{Surveys}} and {{Monographs}}},
   publisher={{American Mathematical Society}},
     address={{Providence, Rhode Island}},
        date={2015},
      volume={209},
        ISBN={978-1-4704-2545-6 978-1-4704-2795-5},
}

\bib{robinsonHypothesisTestingTopological2016}{article}{
      author={Robinson, Andrew},
      author={Turner, Katharine},
       title={Hypothesis {{Testing}} for {{Topological Data Analysis}}},
        date={2016-02},
     journal={arXiv:1310.7467 [cs, math, stat]},
      eprint={1310.7467},
}

\bib{stolzPersistentHomologyTimedependent2017}{article}{
      author={Stolz, Bernadette~J.},
      author={Harrington, Heather~A.},
      author={Porter, Mason~A.},
       title={Persistent homology of time-dependent functional networks
  constructed from coupled time series},
        date={2017-04},
        ISSN={1054-1500},
     journal={Chaos: An Interdisciplinary Journal of Nonlinear Science},
      volume={27},
      number={4},
       pages={047410},
}

\bib{ulmerTopologicalApproach2019}{article}{
      author={Ulmer, M},
      author={Ziegelmeier, Lori},
      author={Topaz, Chad~M},
       title={A topological approach to selecting models of biological
  experiments},
        date={2019-03},
        ISSN={1932-6203},
     journal={PLOS ONE},
      volume={14},
      number={3},
       pages={e0213679},
}

\bib{VICSEK201271}{article}{
      author={Vicsek, Tamás},
      author={Zafeiris, Anna},
       title={Collective motion},
        date={2012},
        ISSN={0370-1573},
     journal={Physics Reports},
      volume={517},
      number={3},
       pages={71\ndash 140},
  url={https://www.sciencedirect.com/science/article/pii/S0370157312000968},
        note={Collective motion},
}

\bib{wackerly02}{book}{
      author={Wackerly, Dennis~D.},
      author={III, William~Mendenhall},
      author={Scheaffer, Richard~L.},
       title={Mathematical statistics with applications},
     edition={sixth edition},
   publisher={Duxbury Advanced Series},
        date={2002},
}

\bib{wadhwaTDAstatsPipelineComputing2018}{article}{
      author={Wadhwa, Raoul~R.},
      author={Williamson, Drew F.~K.},
      author={Dhawan, Andrew},
      author={Scott, James~G.},
       title={{{TDAstats}}: {{R}} pipeline for computing persistent homology in
  topological data analysis},
        date={2018},
     journal={J. Open Source Softw.},
}

\bib{xianCapturingDynamicsTimeVarying2020}{article}{
      author={Xian, Lu},
      author={Adams, Henry},
      author={Topaz, Chad~M.},
      author={Ziegelmeier, Lori},
       title={Capturing {{Dynamics}} of {{Time}}-{{Varying Data}} via
  {{Topology}}},
        date={2020-10},
     journal={arXiv:2010.05780 [cs, math, stat]},
      eprint={2010.05780},
}

\end{biblist}
\end{bibdiv}

\appendix
\FloatBarrier

\end{document}